\documentclass[twocolumn,tighten]{aastex62}

\hypersetup{linkcolor=blue,citecolor=blue,filecolor=green,urlcolor=magenta}

\received{January 1, 9999}
\revised{January 1, 9999}
\accepted{January 1, 9999}

\submitjournal{AAS journal}

\shorttitle{Asteroseismic investigation on KIC 10526294}

\begin{document}

\title{ Asteroseismic investigation on KIC 10526294 to probe convective core overshoot mixing }

\correspondingauthor{Qian-Sheng Zhang}
\email{zqs@ynao.ac.cn}

\author[0000-0003-2449-6226]{Qian-Sheng Zhang}
\affiliation{Yunnan Observatories, Chinese Academy of Sciences, 396 Yangfangwang, Guandu District, Kunming 650216, China}
\affiliation{Center for Astronomical Mega-Science, Chinese Academy of Sciences, 20A Datun Road, Chaoyang District, Beijing 100012, China}
\affiliation{Key Laboratory for the Structure and Evolution of Celestial Objects, Chinese Academy of Sciences, 396 Yangfangwang, Guandu District, Kunming 650216, China}
\affiliation{University of Chinese Academy of Sciences, Beijing 100049, China}
\affiliation{International Centre of Supernovae, Yunnan Key Laboratory, Kunming 650216, P. R. China}

\author[0000-0002-1424-3164]{Yan, Li}
\affiliation{Yunnan Observatories, Chinese Academy of Sciences, 396 Yangfangwang, Guandu District, Kunming 650216, China}
\affiliation{Center for Astronomical Mega-Science, Chinese Academy of Sciences, 20A Datun Road, Chaoyang District, Beijing 100012, China}
\affiliation{Key Laboratory for the Structure and Evolution of Celestial Objects, Chinese Academy of Sciences, 396 Yangfangwang, Guandu District, Kunming 650216, China}
\affiliation{University of Chinese Academy of Sciences, Beijing 100049, China}
\affiliation{International Centre of Supernovae, Yunnan Key Laboratory, Kunming 650216, P. R. China}

\author[0000-0001-6832-4325]{Tao, Wu}
\affiliation{Yunnan Observatories, Chinese Academy of Sciences, 396 Yangfangwang, Guandu District, Kunming 650216, China}
\affiliation{Center for Astronomical Mega-Science, Chinese Academy of Sciences, 20A Datun Road, Chaoyang District, Beijing 100012, China}
\affiliation{Key Laboratory for the Structure and Evolution of Celestial Objects, Chinese Academy of Sciences, 396 Yangfangwang, Guandu District, Kunming 650216, China}
\affiliation{University of Chinese Academy of Sciences, Beijing 100049, China}
\affiliation{International Centre of Supernovae, Yunnan Key Laboratory, Kunming 650216, P. R. China}
\affiliation{Institute of Theoretical Physics, Shanxi University, Taiyuan 030006, China}

\author[0000-0002-7614-1665]{Chen, Jiang}
\affiliation{Max-Planck-Institut f\"{u}r Sonnensystemforschung, Justus-von-Liebig-Weg 3, D-37077 G\"{o}ttingen, Germany}

\begin{abstract}
In the overshoot mixing model with an exponentially decreasing diffusion coefficient, the initial value of the diffusion coefficient plays a crucial role. According to the turbulent convective mixing model, the characteristic length of convection in the convection zone differs from that in the overshoot region, resulting in a rapid decrease of the diffusion coefficient near the convective boundary. To investigate this quick decrease, we conducted an asteroseismic study on the intermediate-mass SPB star KIC 10526294. We generated stellar models with varied input parameters, including the overshoot parameters, and compared the resulting stellar oscillation periods with observations. To mitigate the potential issue arising from large steps in the stellar parameters and stellar age, we employed a comprehensive interpolation scheme for the stellar oscillatory frequencies, considering all stellar parameters and stellar age. Our analysis revealed that the quick decreasing of the diffusion coefficient has discernible effects on the stellar oscillations and a quick decrease with 4 magnitude orders shows the best oscillatory frequencies compared with the observations. This provides weak evidence in support of the prediction made by the turbulent convective mixing model. Furthermore, we examined the residuals of the oscillation periods and discovered a potential association between abundance anomalies in the buoyancy frequency profile and the oscillation-like patterns observed in the residuals.
\end{abstract}

\keywords{asteroseismology --- stars: interiors  --- convection --- stars: individual (KIC 10526294) }

\section{Introduction} \label{SecIntro}

The convective overshoot mixing is an important mixing process in the stellar interior. The overshoot mixing outside a convective core significantly affects stellar evolution by refueling the burning core, prolonging the stellar lifetime, and resulting in a larger helium core. Observations of the extension of the main sequence in the HR diagram confirm the existence of such a mixing process in the stellar interior \citep[see, e.g.,][and the references therein]{Maeder1981}. However, the precise mechanisms through which convective overshooting mixes stellar material remain unclear.

The classical overshoot models are non-local mixing-length theories \citep[e.g.,][]{Shaviv1973,Maeder1975,Bressan1981,Zahn1991} that were developed to determine the penetrating distance, which refers to how far a fluid element can travel across the convective boundary. The penetrating distance represents the extended convective mixing range of the convective core, and it is expected that the convective core should be fully mixed with the extended (overshoot) region. However, the non-local mixing-length models have been found to be self-inconsistent, and the extent of overshooting depends on the model assumption, specifically the adiabatic or radiative stratification in the overshoot region \citep{Renzini1987}. A nearly radiative stratification results in negligible overshooting, while a nearly adiabatic stratification leads to strong overshooting. In models with a nearly adiabatic stratified overshoot region \citep[e.g.,][]{Zahn1991}, the temperature gradient undergoes an abrupt change from adiabatic to radiative. This abrupt variation in the temperature gradient can amplify the amplitude of oscillatory signals in frequencies originating from the base of the solar convection zone \citep{Gough1990}. Analysis of solar p-mode frequencies has indicated a small overshoot length below the base of the solar convection zone \citep{Basu1994,JCD1995}. Furthermore, \citet{JCD2011} found that the required temperature gradient profile is even smoother than the profile predicted by the solar model without overshoot, which cannot be produced by the non-local mixing-length overshoot models.

At present, another widely adopted treatment of the overshoot mixing is to handle it as a diffusion process with an exponential decreasing diffusion coefficient in the overshoot region \citep[e.g.,][]{Freytag1996,Herwig2000,Zhang2012a,Zhang2013,Li2017,ZLCD2019}. This exponentially decreasing behaviour of the diffusion coefficient is based on numerical simulations \citep[e.g.,][]{Freytag1996,Kupka2018} that show exponentially decreasing mean vertical convective velocity in overshoot region, and stellar turbulent convection models \citep[e.g.,][]{Xiong1989,Xiong2002,Zhang2012b,Li2017,Xiong2021} that show exponentially decreasing turbulent kinetic energy and turbulent dissipation. The stellar turbulent convection models are based on hydrodynamic equations, focusing on the turbulent variables (e.g., turbulent kinetic energy, heat flux, temperature variance, turbulent dissipation, and chemical abundance flux) in statistical equilibrium with some closure models of diffusion and dissipation \citep{Xiong1985,Xiong1989,Canuto1993,Canuto1997,Canuto1998,Kupka1999,Xiong2002,Li2007,Zhang2012b,Zhang2013,Li2017,Xiong2021,Kupka2022,Ahlborn2022}. The turbulent convection models are more reasonable and in better agreement than the classical overshoot models when compared with numerical simulations \citep[e.g.,][]{Singh1995,Kupka1999,Kupka2017}. Recently, the properties and structure of the overshoot region predicted by the stellar turbulent convection model \citep{Zhang2012b} have been found to be in good agreement with numerical simulations \citep{Cai2020a,Cai2020b,Cai2020c}. Those models \citep[e.g.,][]{Xiong2002,Zhang2012a,Zhang2012b,Xiong2021} predict a smooth transition of the temperature gradient profile near the convective boundary, which is favored by the investigation of the solar $p$-mode frequencies \citep{JCD2011}.

In the treatment of the overshoot mixing as a diffusion process, the diffusion coefficient is determined by an initial value multiplying an exponentially decreasing factor. In the widely used model of \citet{Herwig2000}, the initial value of the diffusion coefficient is set to be a typical value obtained from the mixing-length theory (MLT) in the convection zone near the convective boundary. On the contrary, \citet{Zhang2013} predicts that the convective characteristic length in the overshoot region differs from that in the convection zone, resulting in a significant decrease of the diffusion coefficient near the convective boundary, followed by an exponential decrease in the bulk of the overshoot region \citep[see, e.g.,][]{Meng2014}. Recently, considering the potential rapid decrease of the diffusion coefficient near the convective boundary, \citet{ZCDL2022} developed a model for overshoot mixing. This model sets the initial value of the diffusion coefficient as the typical value in the convection zone multiplied by a parameter $C$. This model recovers the model of \citet{Herwig2000} when $C=1$, while $C \ll 1$ corresponds to a quick decrease of the diffusion coefficient near the convective boundary. Helioseismic and asteroseismic investigations \citep{ZCDL2022} on solar models and 13 Kepler solar-like stars have not placed any constraints on the parameter $C$. In this work, we plan to conduct asteroseismic investigations on stars exhibiting $g$-mode oscillations to explore the possibility of a rapid decrease.

Asteroseismology, which studies wave oscillations throughout the stellar interior, is a powerful tool for probing stellar structure and investigating physical processes within stars. Two types of asteroseismic investigations are commonly conducted to explore convective core overshoot. The first type investigates the ratios of small to large frequency separations in stellar $p$-mode oscillations, while the second type examines stellar $g$-mode and mixed-mode oscillations, as both are sensitive to the structure of the stellar core. \citet{Deheuvels2011} provided detailed modeling of avoided crossings to constrain the stellar properties of HD 49385. \citet{Yang2015} analyzed the frequency separation ratios to identify a significant overshoot distance in KIC 2837475. \citet{Deheuvels2016} studied the ratio of small to large frequency separations in 24 \textsl{CoRoT} and \textsl{Kepler} low-mass stars and observed a trend of increasing overshoot region size with stellar mass. \citet{Moravveji2015} modeled the oscillations of the slowly pulsating B-type (SPB) star KIC 10526294 and found that the diffusion overshoot mixing model of \citet{Herwig2000} provided a better fit to the observed oscillation periods than the classical non-local mixing-length overshoot model. Similar results were obtained by \citet{Moravveji2016} in their asteroseismic investigation of another intermediate-rotating SPB star, KIC 7760680, and they suggested that KIC 10526294 and KIC 7760680 are ideal targets for testing theories of convection. \citet{Wu2019} examined the oscillations of the SPB star HD 50230 and constrained the stellar parameters and overshoot mixing. \citet{Wu2020} also investigated the SPB star KIC 8324482 and identified a weak convective core overshoot mixing. \citet{Michielsen2019} demonstrated the capability of asteroseismology in probing thermal stratification in the overshoot region. Recently, \citet{Michielsen2021} analyzed the SPB star KIC 7760680 and found that a radiative stratification in the overshoot region is more favorable than an adiabatic-dominated stratification. \citet{Noll2021} investigated KIC 10273246 and traced the evolution of mixed-mode frequencies to constrain core overshoot. They found that the data could not distinguish between exponential decreasing diffusion and constant diffusion in the overshoot region. \citet{Pedersen2021} modeled 26 SPB stars with different internal mixing process patterns and investigated them using asteroseismology. They found that internal mixing profiles in SPB stars exhibit radial stratification. No fixed pattern emerged as the best fit for all target stars, and exponential decreasing diffusion of overshoot did not yield better results than constant diffusion of overshoot.

In this paper, we conduct an asteroseismic investigation of the SPB star KIC 10526294 to explore core overshoot, particularly focusing on the rapid decrease of the diffusion coefficient near the convective boundary. Section \ref{SecKIC} provides a brief introduction to our target, KIC 10526294. In Section \ref{SecOVM}, we describe the overshoot mixing model adopted in detail. The input physics used in the calculations of the stellar models are outlined in Section \ref{SecInput}. The numerical results are presented in Sections \ref{SecGrids} and \ref{SecResult}. We discuss the obtained results in Section \ref{SecDis}. Finally, we summarize our conclusions in Section \ref{SecConclusion}.

\section{About KIC 10526294} \label{SecKIC}

KIC 10526294 has been classified as an SPB star, and a detailed analysis of its 4-year light curve revealed an oscillatory period range of approximately $1-2$ days, including 19 individual eigenfrequencies, with a mean period spacing of about 0.06 day \citep{Papics2014}. The analysis of its spectra provided the atmospheric parameters of the star: an effective temperature of $T_{\rm eff}=11550\pm500$ K, a surface gravitational acceleration of $\log g=4.1\pm0.2$ (in c.g.s. units), and a metallicity of $Z = 0.016 ^{+0.013} _{-0.007}$ \citep{Papics2014}. The rotation period of KIC 10526294 is estimated to be approximately 188 days based on the average splitting of stellar oscillation frequencies \citep{Papics2014}. The internal rotation profile of KIC 10526294 has been analyzed by \citet{Triana2015} using frequency splitting \citep{Aerts2010}. They found that the average rotation rate near the core overshoot region is approximately 163$\pm$89 nHz, and a mild counter-rotating region in the envelope towards the surface is preferred when assuming a smooth profile of rotation rate in the stellar interior.

Asteroseismic investigations based on standard stellar models \citep{Papics2014} have indicated that KIC 10526294 is a young star with a stellar mass of about 3.2 solar masses, and it has an upper limit for core overshoot of $\alpha_{\rm ov}\leq0.15$ for classical fully mixing or $f_{\rm ov}\leq0.015$ for the diffusion overshoot model of \citet{Herwig2000}. Further comprehensive asteroseismic investigations on stellar models of KIC 10526294 have been conducted by \citet{Moravveji2015}, and subsequent investigations based on a statistical approach have been performed by \citep{Aerts2018} and \citet{Pedersen2021}. \citet{Moravveji2015} examined stellar models on a denser grid, considering an additional global mixing, testing different compositions, and comparing classical fully overshoot mixing with the diffusion overshoot model of \citet{Herwig2000}. They found that the effects of the additional global mixing are significant in fitting an oscillation-like pattern of the period spacing. The $\chi^2$ value of the best stellar model with the diffusion overshoot model is less than half of the $\chi^2$ value of the best stellar model with classical overshoot fully mixing, indicating that the diffusion overshoot model is more reasonable than the classical overshoot models. All these analyses of KIC 10526294 highlight it as an excellent target for studying overshoot and testing the theory of diffusion overshoot mixing. Therefore, in this work, we continue to model KIC 10526294 using the diffusion overshoot mixing approach and aim to investigate the possible rapid variation of the diffusion coefficient near the convective boundary. We hope that such an analysis can provide further insights into the current developments of stellar convection theory.

\section{Convective overshoot mixing model} \label{SecOVM}

\begin{figure}
	\includegraphics[width=\columnwidth]{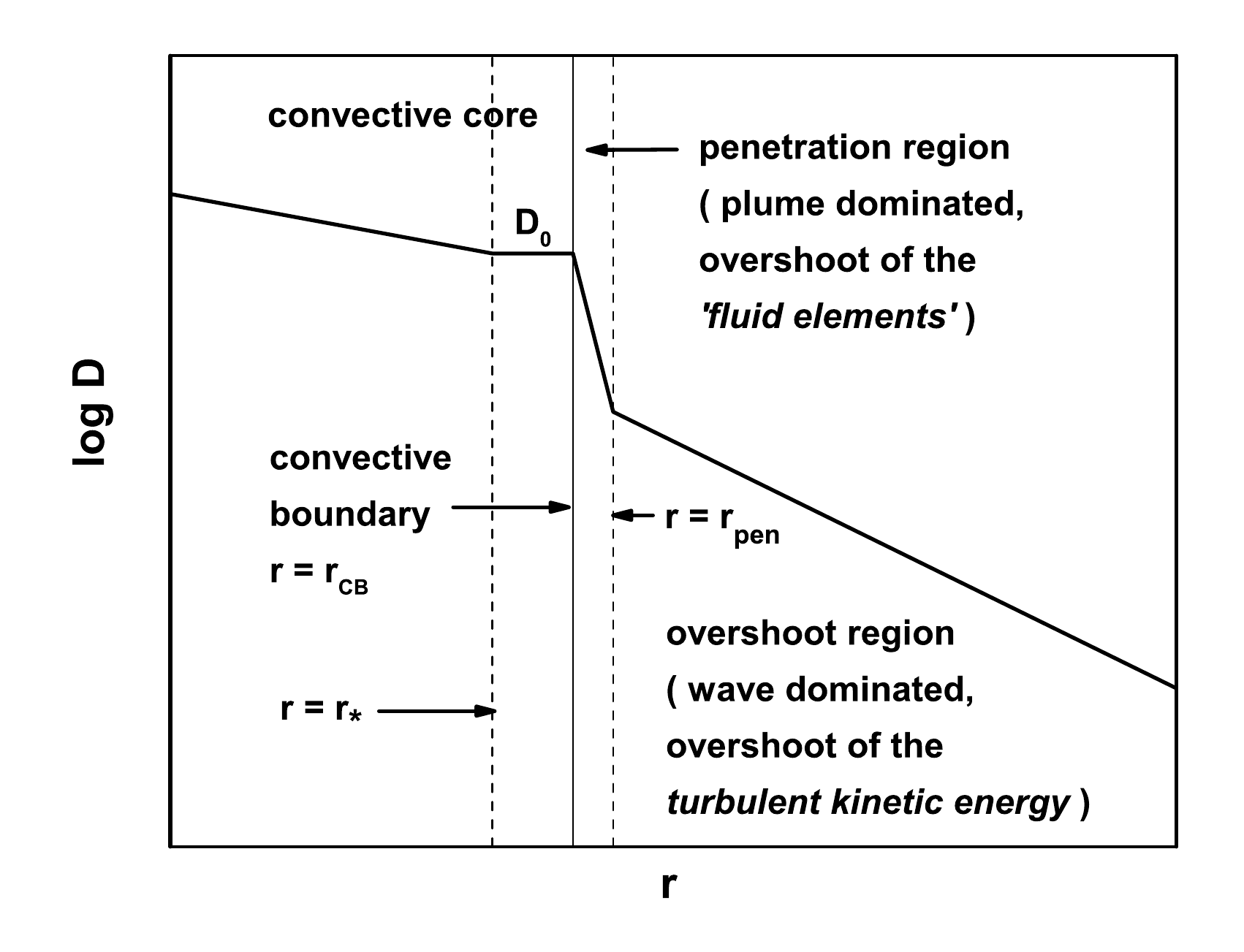}
    \caption{A sketch of the diffusion coefficient of the convective/overshoot mixing in the stellar interior. $D_0$ is the typical diffusion coefficient in the convective core near the convective boundary, calculated using $D_0=u(r_*)l/3$, where $l$ is the mixing length and $u$ is the convective speed calculated determined by the MLT. The point $r_*$ is chosen within the convective core near the boundary. The value of $D_0$ is used for $r_* < r < r_{\rm CB}$ to avoid the local property issues associated with MLT. In the penetration region, which corresponds to the overshoot of \textsl{fluid elements}, a rapid decrease in the diffusion coefficient occurs. This region spans from $r_{\rm CB}$ to $r_{\rm pen}$, where the temperature gradient transitions from nearly adiabatic to nearly radiative. In the wave-dominated bulk overshoot region ($r > r_{\rm pen}$), the diffusion coefficient $D$ exponentially decreases. It is important to note that the \textsl{penetration region} refers to the overshoot of \textsl{fluid elements}, while the \textsl{overshoot region} pertains to the overshoot of \textsl{turbulent kinetic energy}. In this study, the penetration region is disregarded, and the rapid decrease is assumed to occur precisely at the convective boundary. This simplification is made because the depth of the penetration region cannot be directly determined from first principles and is typically small according to turbulent models \citep[e.g.,][]{Zhang2012b,Li2017}. This approximation is acceptable since the thin penetration layer can be well mixed even if $D$ is multiplied by a small decreasing factor $C$ (e.g., $10^{-6}$). Consequently, $D$ exhibits a sudden decrease at the convective boundary.}
    \label{Dovsketch}
\end{figure}

The adopted model for convective core overshoot mixing is based on an exponentially decreasing diffusion coefficient, as developed by \citet{ZCDL2022}. The model is described by the following equation:
\begin{eqnarray} \label{OVMZCDL2022}
D_{\rm ov}=CD_0(\frac{P}{P_{\rm CB}})^{\theta},
\end{eqnarray}%
where $D_{\rm ov}$ represents the diffusion coefficient for overshoot mixing, $P$ denotes pressure, $P_{\rm CB}$ represents the pressure at the boundary of the convective core, $D_0$ corresponds to the typical diffusion coefficient near the boundary of the convective core, and $C$ and $\theta$ are parameters that characterize the properties of the model. The specific details of these parameters will be introduced later. To determine $D_0$, the typical value of the diffusion coefficient near the boundary of the convective core, the MLT is employed. In this approach, $D_0$ is taken as $D(r_*)$, where $r_*$ is a location inside the convective core chosen to avoid the local properties of MLT. In this case, $r_* = r_{CB} - d$, with $r_{CB}$ representing the convective boundary and $d$ denoting a nonzero distance, which is set to 0.1 $H_P$ by default. A graphical representation of the diffusion coefficient is presented in Figure \ref{Dovsketch}.

The meanings of the model have been discussed in detail in \citet{ZCDL2022}. Here, we summarize the main properties of the model. The parameter $\theta$ represents the factor responsible for the exponential decrease in $D$, primarily determining the effective length of the overshoot mixing. The parameter $C$ describes the potential quick variation of the diffusion coefficient near the convective boundary. When $C\ll1$, it corresponds to a significant rapid decrease in $D$. This rapid decrease is assumed to occur within a very thin layer, and we do not trace the precise details of this rapid decline. Consequently, combining Eq.(\ref{OVMZCDL2022}) with $D$ in the convection zone leads to a discontinuity at the boundary. $C$ also secondarily determines the effective length of the overshoot mixing. The model reverts to the widely used formulation of \citet{Herwig2000} when $C=1$ and $\theta=2/f_{\rm ov}$. Here, $C=1$ signifies the absence of a quick variation in $D$ near the boundary, which is an implicit assumption in the formula proposed by \citet{Herwig2000}.

\section{Input physics in calculations of stellar models} \label{SecInput}

Stellar models and oscillations are calculated using the YNEV code \citep{YNEVZ15}. The AGSS09 \citep{AGSS09} solar composition, revised for Ne \citep{Young2018}, is adopted for elements heavier than He in the stellar models. This composition is essentially the same as the recently revised solar composition proposed by \citet{Asplund2021}. The opacity in the stellar interior is interpolated from the OPAL opacity tables \citep{Iglesias1996} based on the adopted composition. The low-temperature opacity tables \citep{Ferguson2005} are smoothly connected to the OPAL tables. The OPAL equation-of-state tables \citep{Rogers2002} are used to interpolate pressure $P$, its derivatives with respect to density and temperature $(\partial {\rm ln} P / \partial {\rm ln} \rho )_T$, $(\partial {\rm ln} P / \partial {\rm ln} T )_{\rho}$, and the adiabatic temperature gradient $\nabla_{\rm ad}=(\partial {\rm ln} T / \partial {\rm ln} P)_S$. Other quantities such as the adiabatic compression index $\Gamma_1$ and specific capacities $c_V$, $c_P$ are obtained using thermodynamic relations. Nuclear reaction rates are based on NACRE \citep{Angulo1999} and are enhanced by weak screening \citep{Salpeter1954}. The temperature gradient in the convective core/zones is calculated using the MLT with $\alpha=1.8$, which is calibrated in the solar model. Since the envelope of a main-sequence 3$M{\odot}$ star is radiative-dominated, the effects of varying $\alpha$ are not significant. The convective heat flux in the overshoot region is ignored. Apart from the convective overshoot mixing and a global diffusion with $ {\rm log} D_{\rm ext} =1.75 $, no other non-standard physical processes such as microscopic diffusion, mass loss, or accretion are taken into account.

In order to reduce the number of stellar models, we adopt a universal helium enhancement law given by $Y=Y_p+(\Delta Y/\Delta Z)Z$ for the initial stellar abundances. Here, $Y_p$ represents the primordial helium abundance, and we adopt the value $Y_p=0.2485$ based on the Big Bang Nucleosynthesis (BBN) result \citep{Cyburt2008}. The helium enhancement ratio $\Delta Y/\Delta Z$ is set to 2.0 by comparing the enhancement law with the standard chemical mixture observed in OB stars in the solar neighborhood ($Y=0.276$, $Z=0.014$) \citep{Nieva2012}. However, there is evidence suggesting that the linear helium enhancement law might be too simplistic, and the helium enhancement ratio could vary over a wide range \citep[see, e.g., ][and references therein]{Verma2019}. In asteroseismic investigations of SPB stars, an additional global diffusion process has been found to be widely present in order to explain the observed period spacing patterns \citep[see, e.g., ][]{Moravveji2015,Moravveji2016,Wu2019,Wu2020,Pedersen2021}. For KIC 10526294, we adopt a global extra diffusion with $ {\rm log} D_{\rm ext} =1.75 $ in our calculations, based on the results obtained by \citet{Moravveji2015}. We will discuss the effects of varying these two constraints, namely the reduced freedom in the initial stellar abundances and the fixed value of $ {\rm log} D_{\rm ext}$.

All stellar models in this work evolve from the pre-main sequence (PMS) stage with a central temperature of $T_{\rm C}=10^5$ K to the end of the main sequence, defined as the point when the central hydrogen abundance decreases to 0.01. The stellar models typically consist of around 2000 mesh points. The time step is controlled to ensure that the variation of the hydrogen abundance in the stellar core is less than 0.005 within a single time step. As a result, there are approximately 200 models along the evolutionary track.

\section{Coarse and fine grids} \label{SecGrids}

We have calculated 11,988 stellar evolutionary tracks using a coarse grid with varied stellar parameters covering various ranges, as listed in Table \ref{tablegrids}. This coarse grid, with relatively low resolutions of the stellar parameters, provides an initial exploration of the parameter space where the computed stellar models exhibit good agreement with observations. Subsequently, a second step involves calculating additional models with a finer grid based on the preliminary parameter space to search for the best-fitting models.

\begin{table}
\centering
\caption{ Stellar parameters and their ranges for the grids of stellar models. }\label{tablegrids}
\begin{tabular}{lcccc}
\hline\noalign{\smallskip}
                                      & from       & to       & step     & points  \\
\hline
        coarse grid                   &            &          &          & 11988   \\
        stellar mass $M$ ($M_\odot$)  & 2.70       & 3.60     & 0.05     & 37      \\
        metallicity $Z$               & 0.008      & 0.040    & 0.008    &  9      \\
       $ \log C $                     & -6         & 0        & 2        &  4      \\
       $ \log \theta $                & 1.4        & 3.0      & 0.2      &  9      \\
\hline
          fine grid                   &            &          &          & 233393  \\
        $ S=M+16Z $                  & 3.40       & 3.70     & 0.01     & 31      \\
            $Z$                       & 0.008      & 0.040    & 0.002    & 17      \\
       $ \log C $                     & -6         & 0        & 0.5      & 13      \\
       $ \log \theta $                & 1.30       & 2.30     & 0.02     & $\sim$20\\
\hline
\end{tabular}
\end{table}

\begin{figure}
    \centering
	\includegraphics[width=\columnwidth]{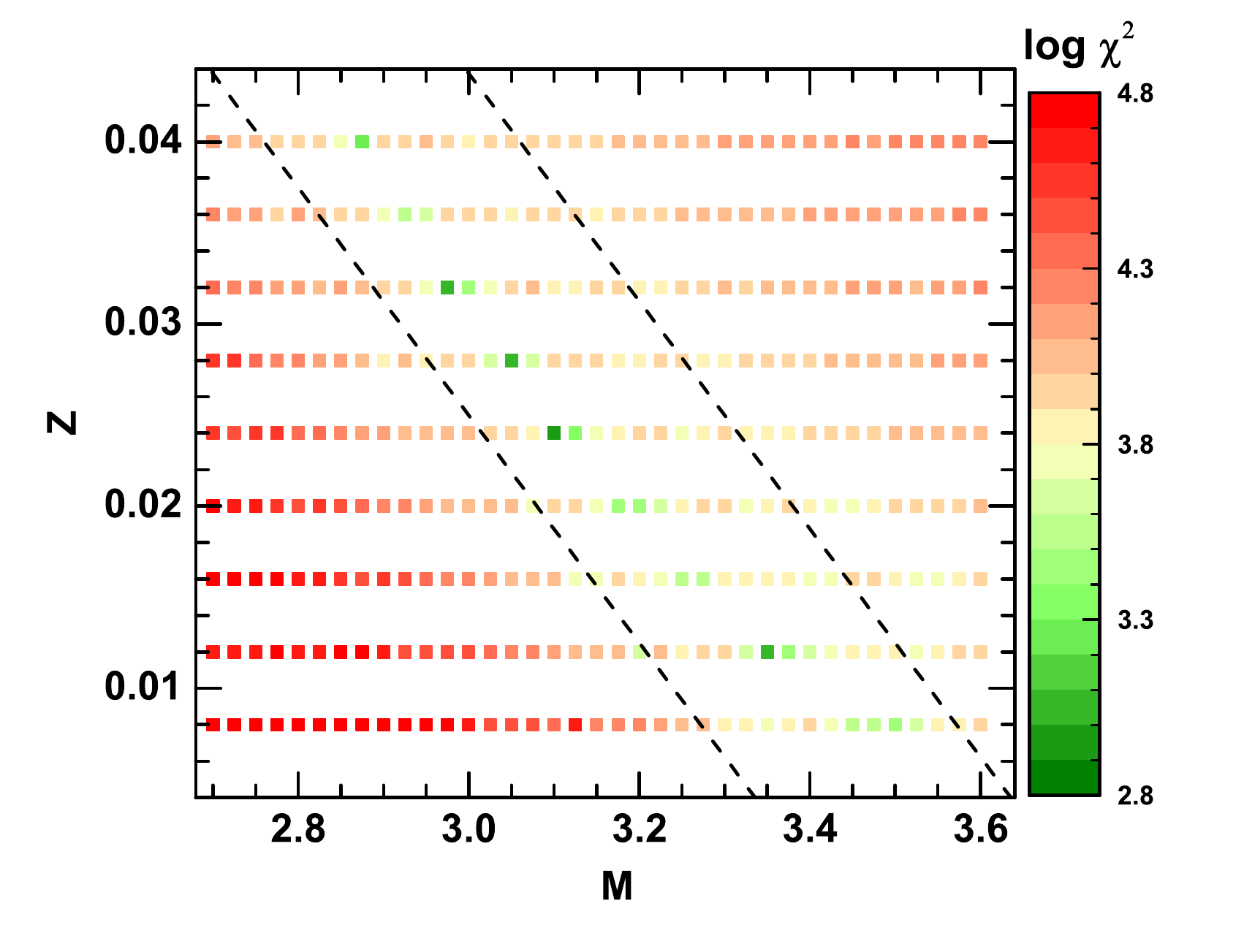}
	\includegraphics[width=\columnwidth]{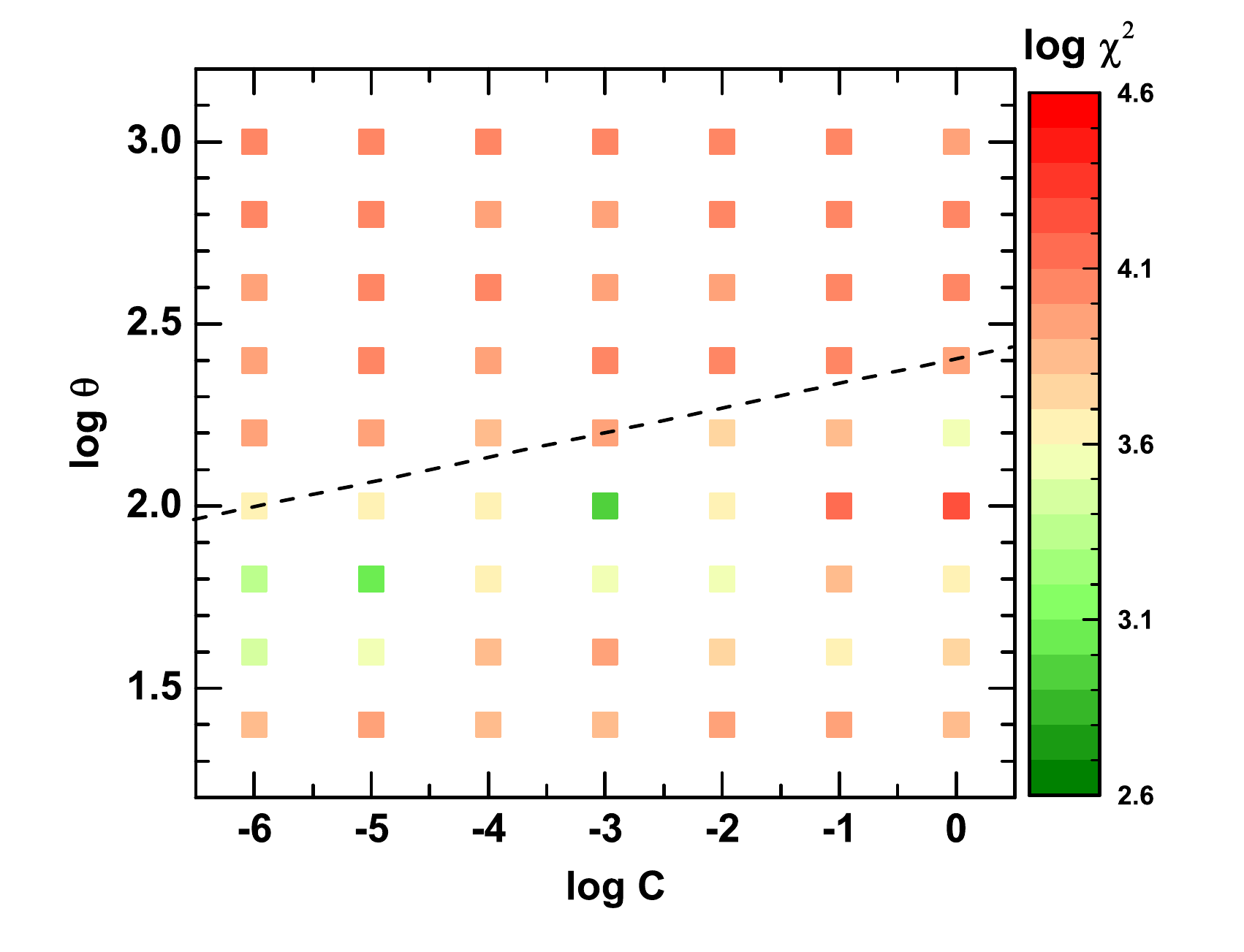}
    \caption{The top panel: $\log \chi^2$ vs. $M$, $Z$ for all points of $M$ and $Z$ on the coarse grid, where $\chi^2$ is the minimum of $\chi^2$ for all stellar models with given $M$, $Z$ and varied $\log C$, $\log \theta$ and $\tau$. The bottom panel: $\log \chi^2$ vs. $\log C$, $\log \theta$ for all points of $\log C$ and $\log \theta$ on the coarse grid, where $\chi^2$ is now the minimum of $\chi^2$ for all stellar models with given $\log C$ and $\log \theta$ and varied $M$, $Z$ and $\tau$. }
    \label{coarse}
\end{figure}

A stellar evolutionary track is determined by the stellar parameters at each mesh point on the grid. Along each track, the oscillation frequencies depend on the stellar age $t$. We evaluate the stellar models using the $\chi^2$ statistic defined as follows:
\begin{eqnarray} \label{chi2def}
{\chi ^2} = \frac{1}{{{N_{{\rm{obs}}}}}}\sum\limits_{k = 1}^{{N_{{\rm{obs}}}}} {{{\left(\frac{{{P_k} - {P_{k,{\rm{obs}}}}}}{{\sigma {P_{k,{\rm{obs}}}}}}\right)}^2}},
\end{eqnarray}
where $N_{\rm obs} = 19$ represents the number of observed stellar oscillatory periods. The observed periods and their uncertainties are obtained from \citet{Moravveji2015}. As a stellar evolutionary track is determined by the stellar parameters, and the interior structure of a star on the track is determined by its age, the value of $\chi^2$ becomes a function of ($\log C$, $\log \theta$, $M$, $Z$, $t$). Along a given evolutionary track, we can identify the best-fitting model at a particular age, denoted as $t_m$, which reproduces the observations with the minimum value of $\chi^2$ ($\chi^2_m$). Both $\chi^2_m$ and $t_m$ depend on the stellar parameters ($\log C$, $\log \theta$, $M$, $Z$).

No significant dependence of $\chi^2_{\rm m}$ on individual parameters $C$, $M$, or $Z$ has been observed in the coarse grid. However, as depicted in the top panel of Figure \ref{coarse}, $\chi^2$ exhibits a dependence on a linear combination of $M$ and $Z$. The optimal range for $M$ and $Z$ is indicated by the two dashed lines, defined as $3.4 \leq S=M/M_{\odot}+16Z \leq 3.7$. The dependence of $\chi^2$ on $\log C$ and $\log \theta$ is shown in the bottom panel of Figure \ref{coarse}. Mesh points that lie below or near the dashed line, represented by $\log \theta = 0.05 \log C + 2.3 $, generally yield lower $\chi^2$ values, suggesting an optimal range for these two parameters. However, two mesh points, specifically ($\log \theta = 1.6$, $\log C = -1 $) and ($\log \theta = 1.8$, $\log C = 0 $), show local minima of $\chi^2$. Considering the sparsity of the parameter space in the coarse grid, we decided to include all points below the dashed line as the optimal region. This approach allows us to further constrain the models using a fine grid generated based on the two criteria outlined in Figure \ref{coarse}. To recap, the optimal parameter space is summarized as $3.40 \leq S \leq 3.70$ and $1.3 \leq \log \theta \leq 0.05 \log C + 2.3 $. We computed a total of 233,393 stellar evolutionary tracks in the fine grid, covering the specified parameter ranges with small steps for each stellar parameter, as listed in Table \ref{tablegrids}. In the fine grid, $S$, $Z$, $ \log C $, and $ \log \theta $ are evenly spaced. Since $S$ is a function of stellar mass, the step size for $M$ is not fixed, and the range of $M$ in the fine grid is from 2.76 to 3.572. Subsequently, we constrained the range of each stellar parameter ($M$, $Z$, $\log \theta$, $\log C$) and some global variables (age, effective temperature, radius, central hydrogen abundance, etc.) of the stellar models by analyzing the distribution of $\chi^2_{\rm m}$ for each of these parameters.

\section{Results} \label{SecResult}

\subsection{Suggested ranges of stellar parameters and global variables} \label{SecResult1}

The minimum value of $\chi^2_{\rm m}$ among all the evolutionary tracks in the fine grid is $\chi^2_{\rm best}=495$. Using an interpolation of oscillation frequencies (to be described later) with respect to the stellar parameters $\log \theta$, $M$, $Z$, and stellar age, we identified the best model with $\chi^2_{\rm best}=442$. Following the approach of \citet{ZCDL2022}, where it is assumed that $\chi^2$ values outside the suggested range of stellar parameters should be significantly larger than those within the suggested range, we defined the suggested range as $\chi^2_{\rm m}<2\chi^2_{\rm best}$. Therefore, the range of stellar parameters satisfying $\log \chi^2_{\rm m} < 2.95$ is considered the suggested range.

\begin{figure*}
    \centering
	\includegraphics[width=0.66\columnwidth]{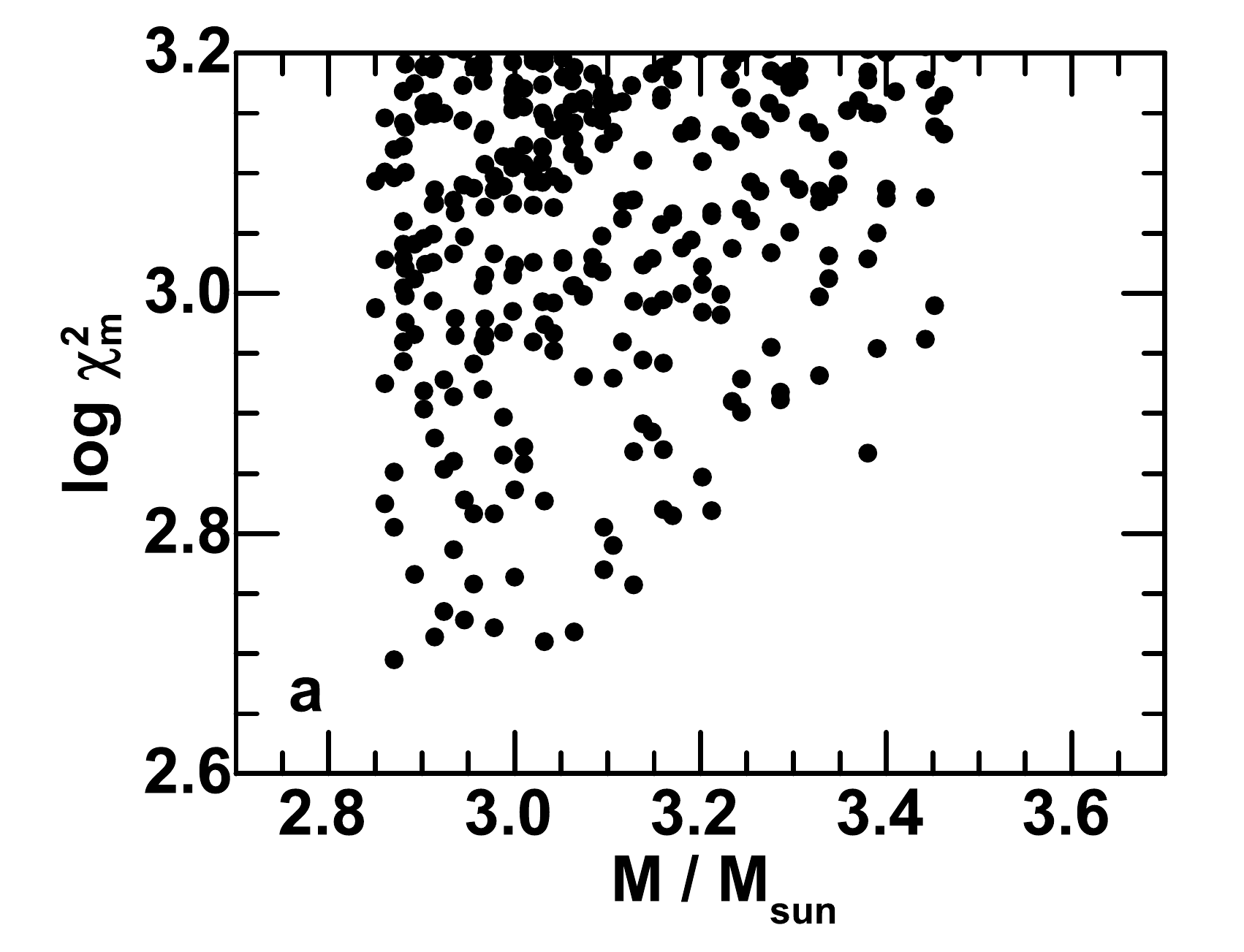}
	\includegraphics[width=0.66\columnwidth]{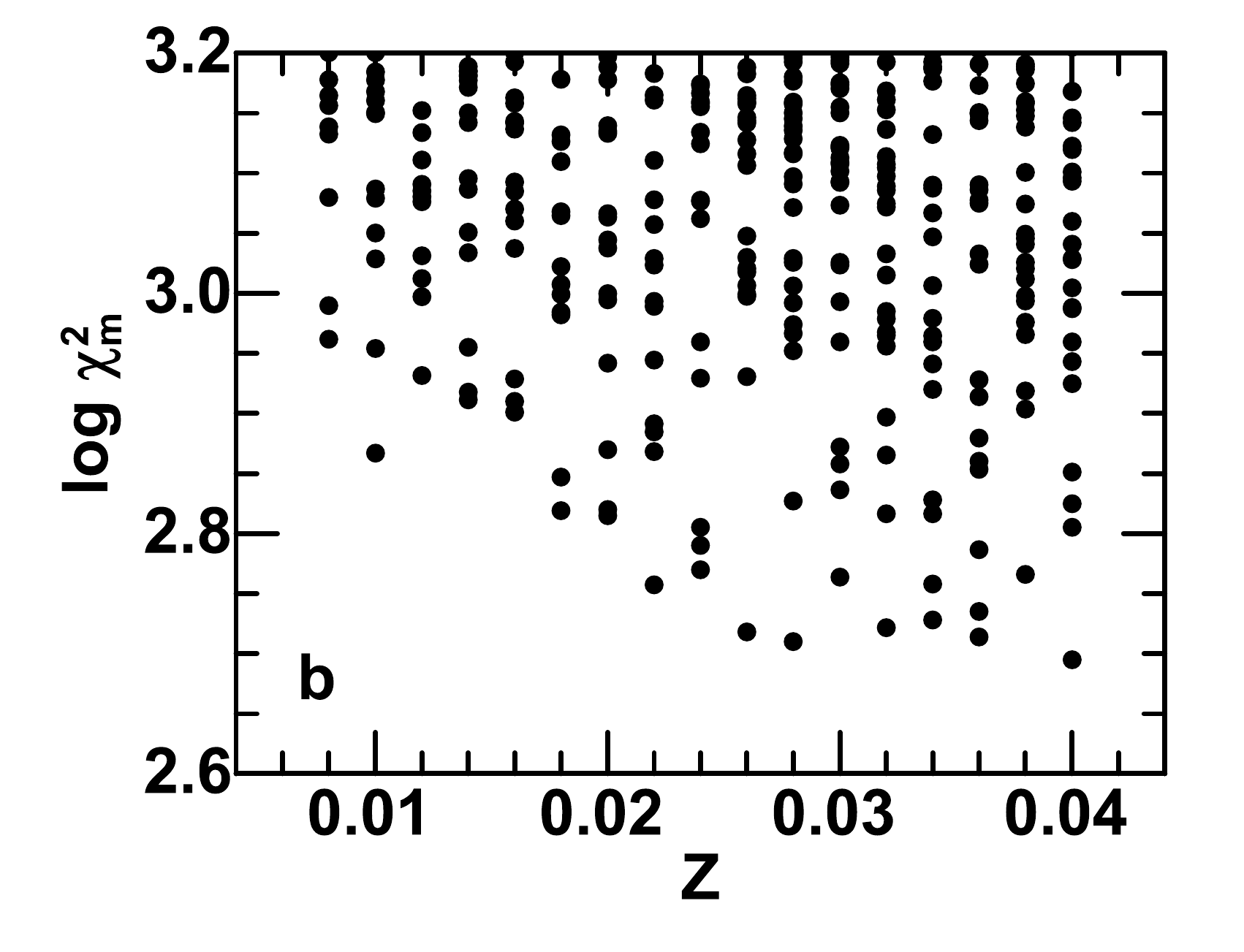}
	\includegraphics[width=0.66\columnwidth]{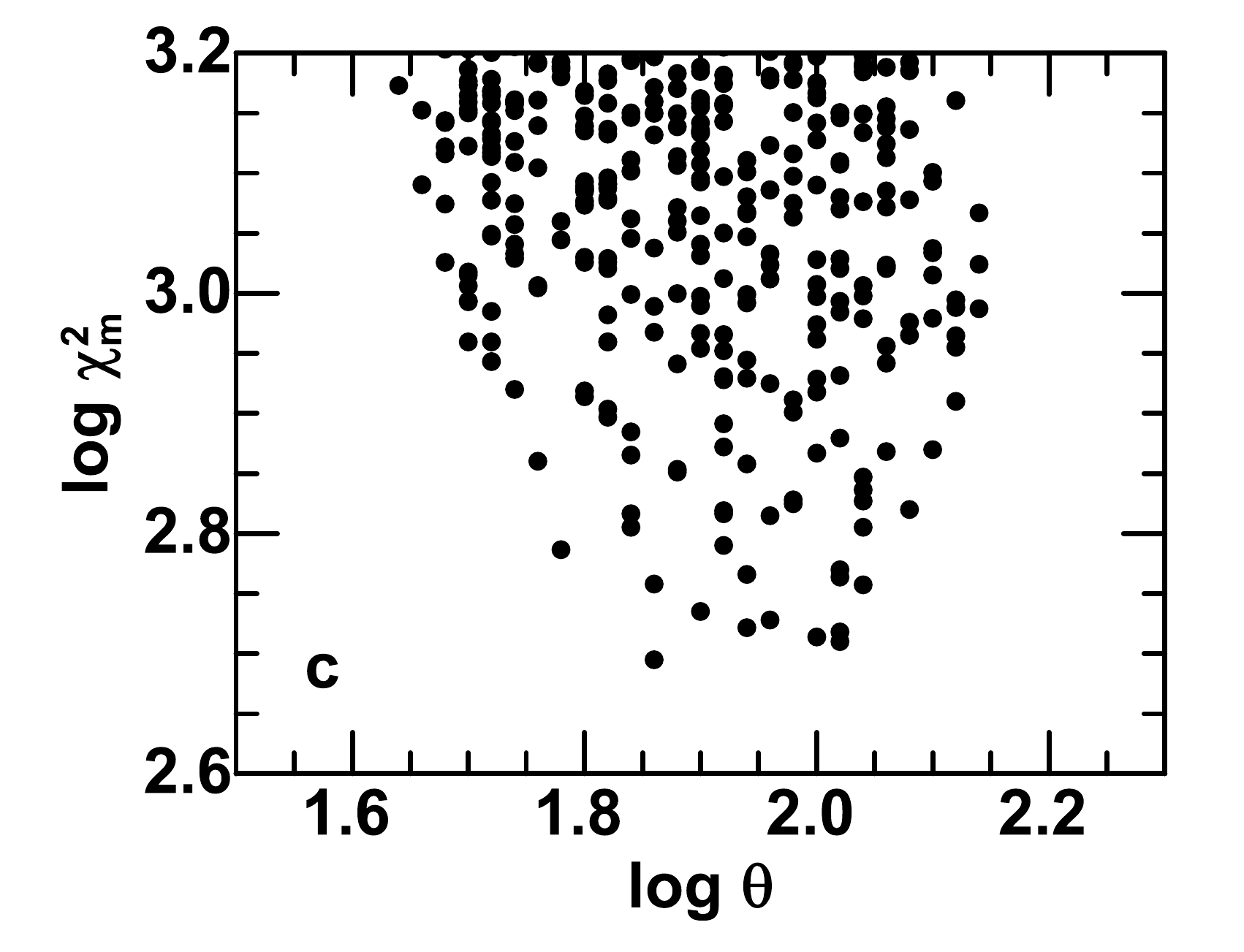}
	\includegraphics[width=0.66\columnwidth]{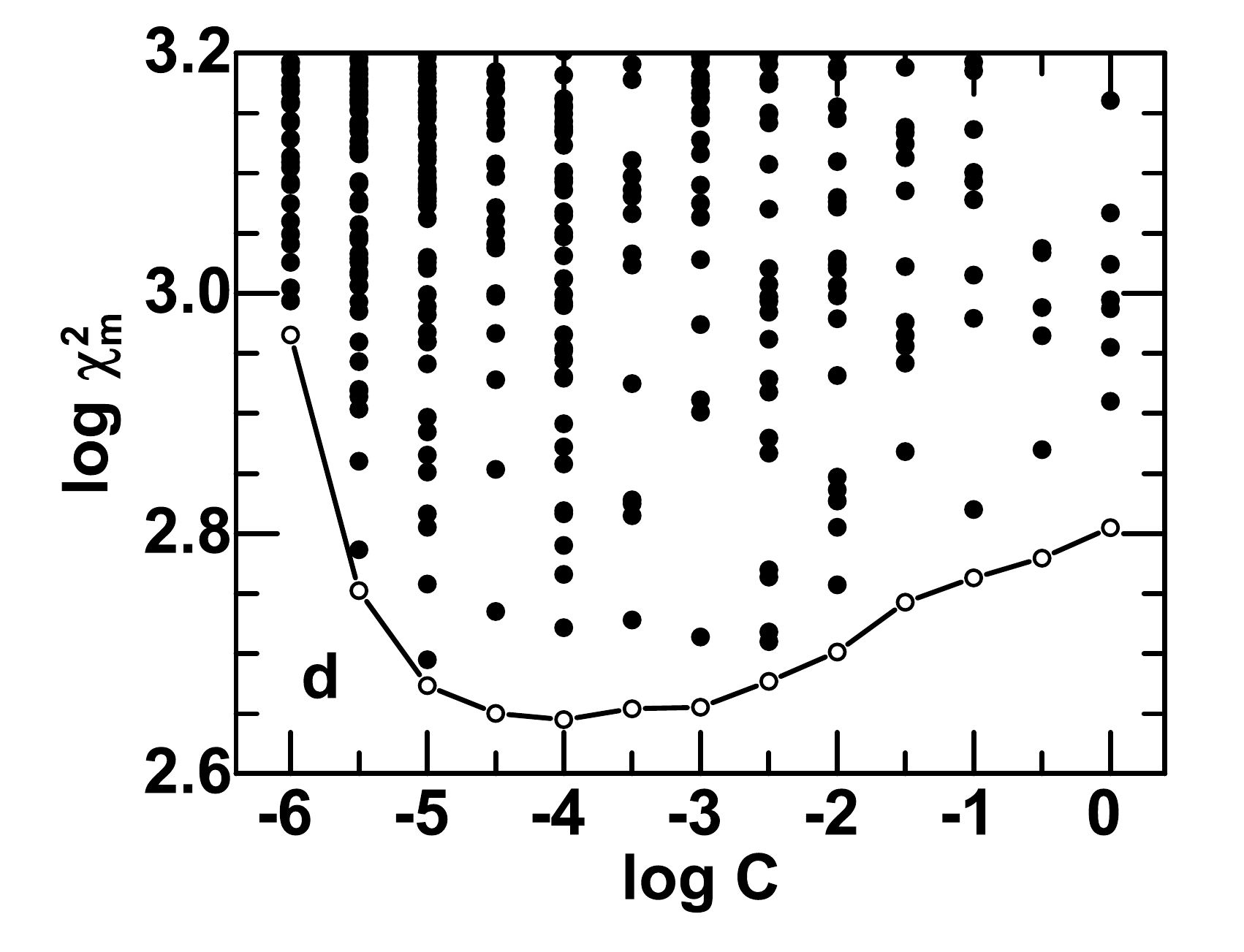}
	\includegraphics[width=0.66\columnwidth]{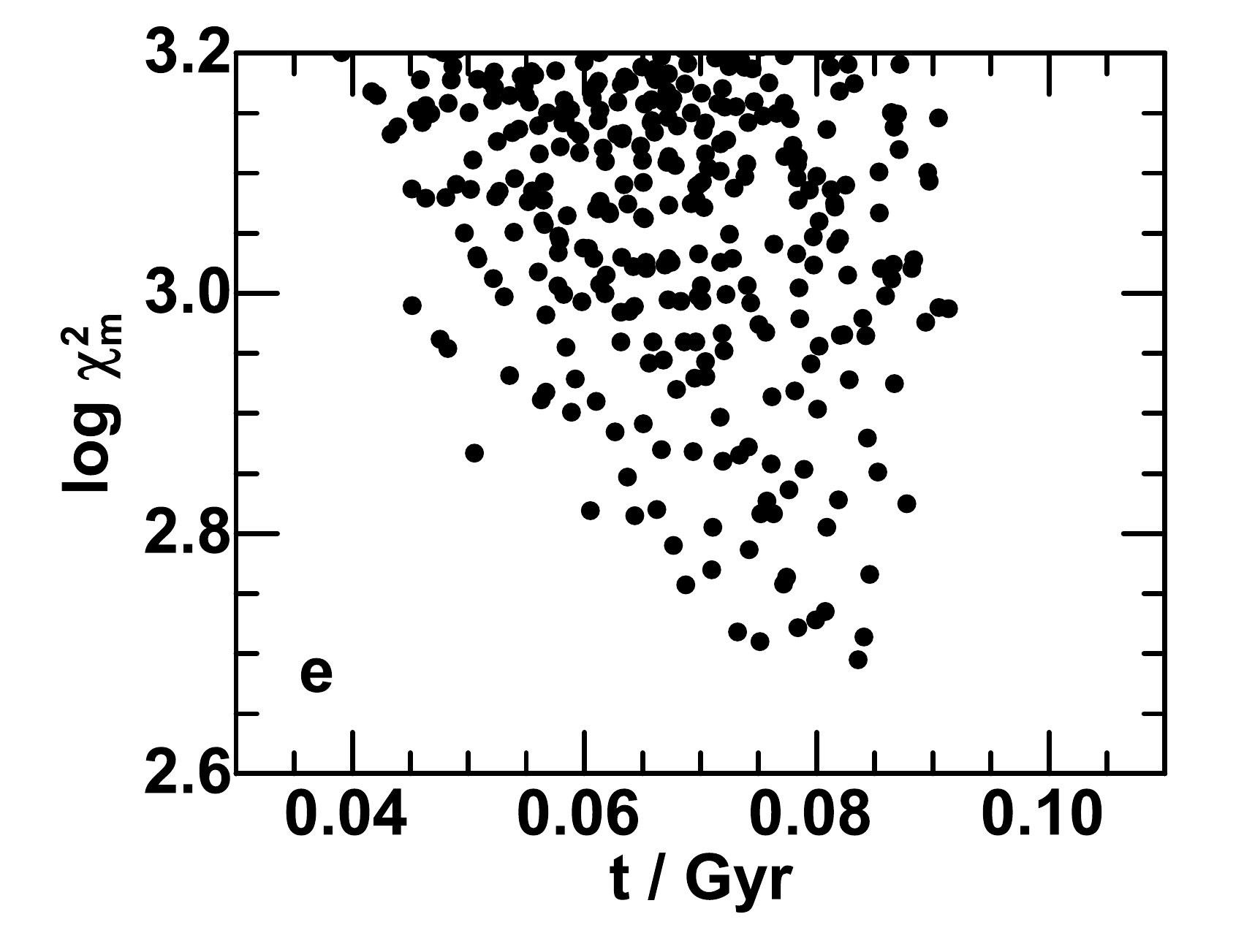}
	\includegraphics[width=0.66\columnwidth]{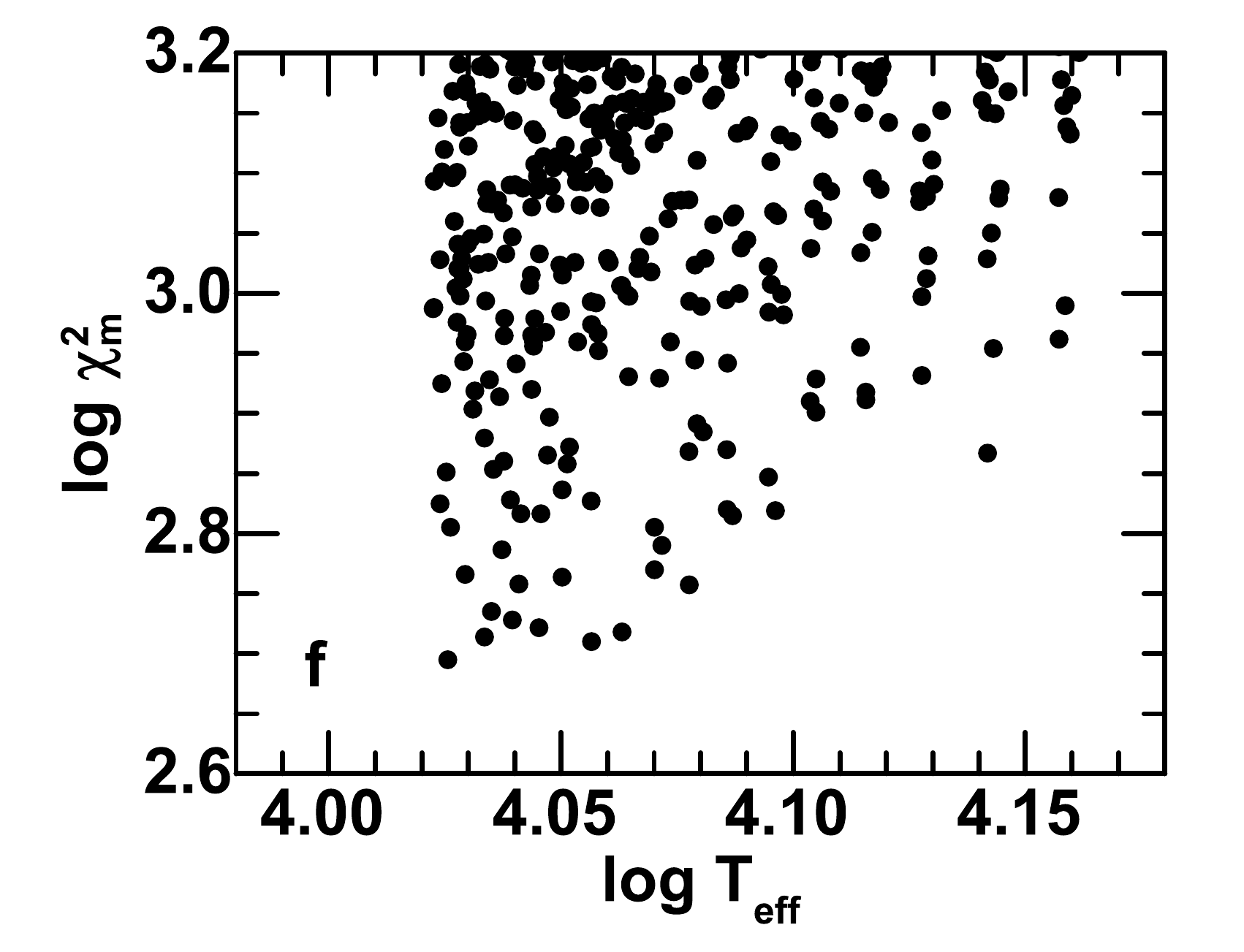}
	\includegraphics[width=0.66\columnwidth]{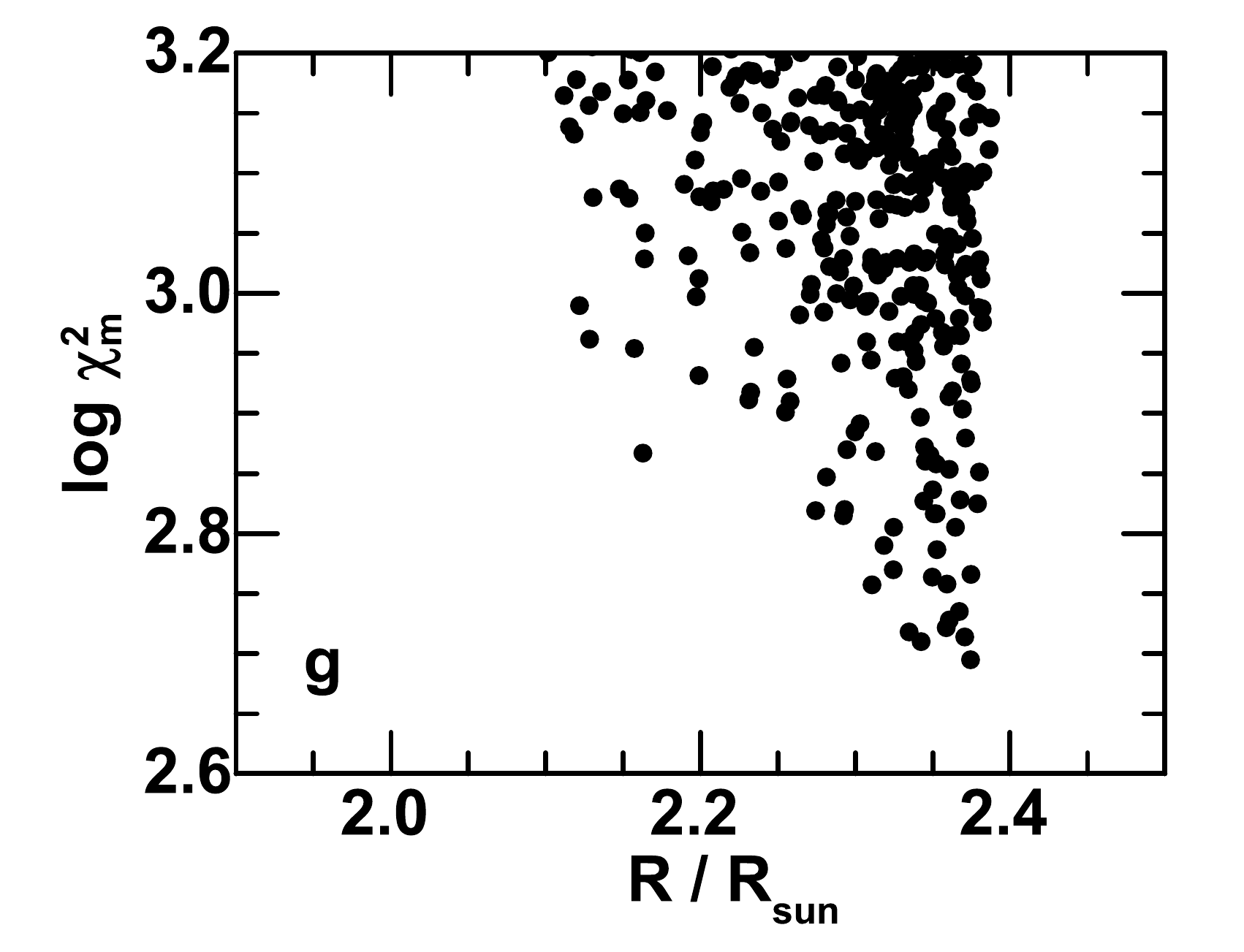}
	\includegraphics[width=0.66\columnwidth]{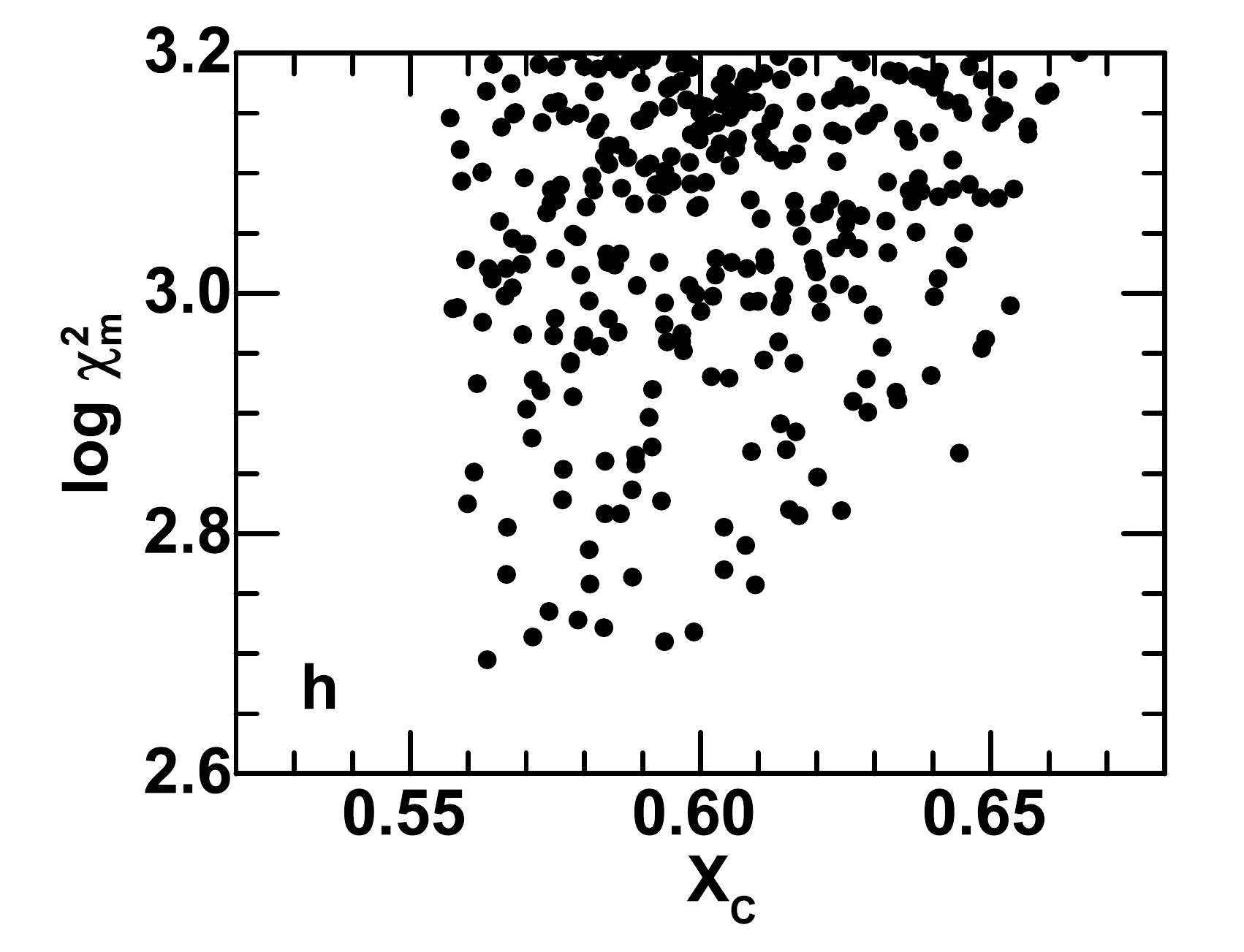}
	\includegraphics[width=0.66\columnwidth]{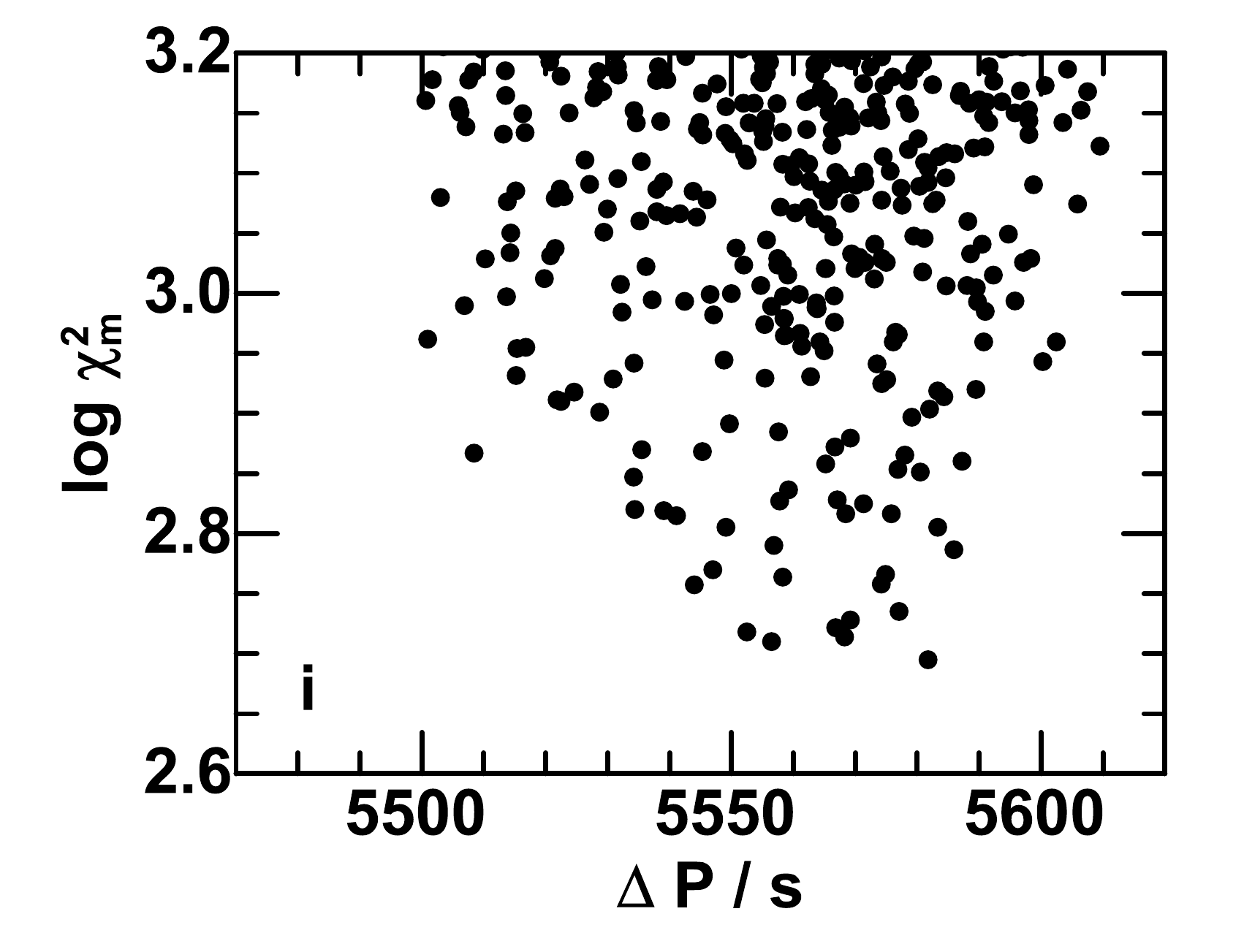}
	\includegraphics[width=0.66\columnwidth]{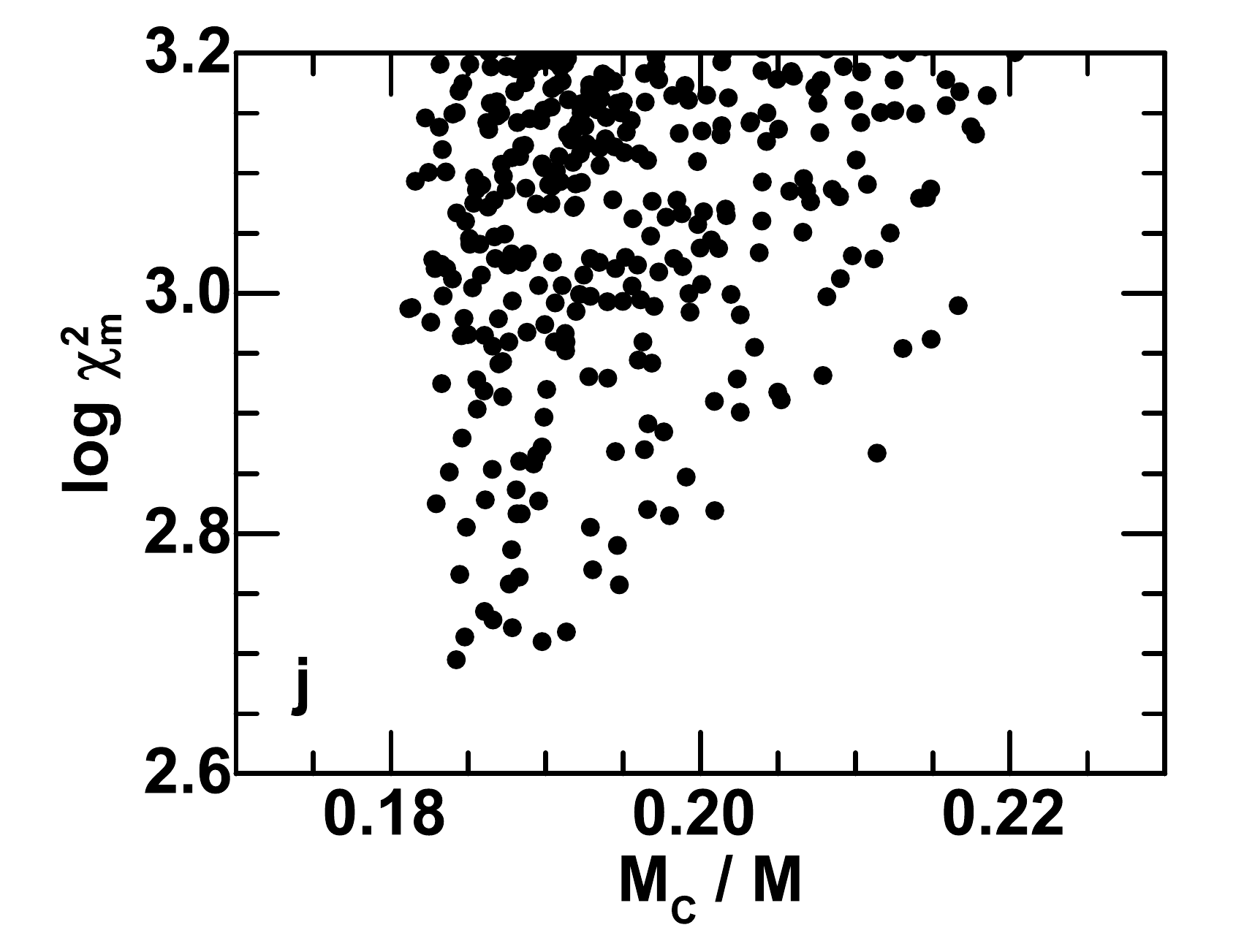}
	\includegraphics[width=0.66\columnwidth]{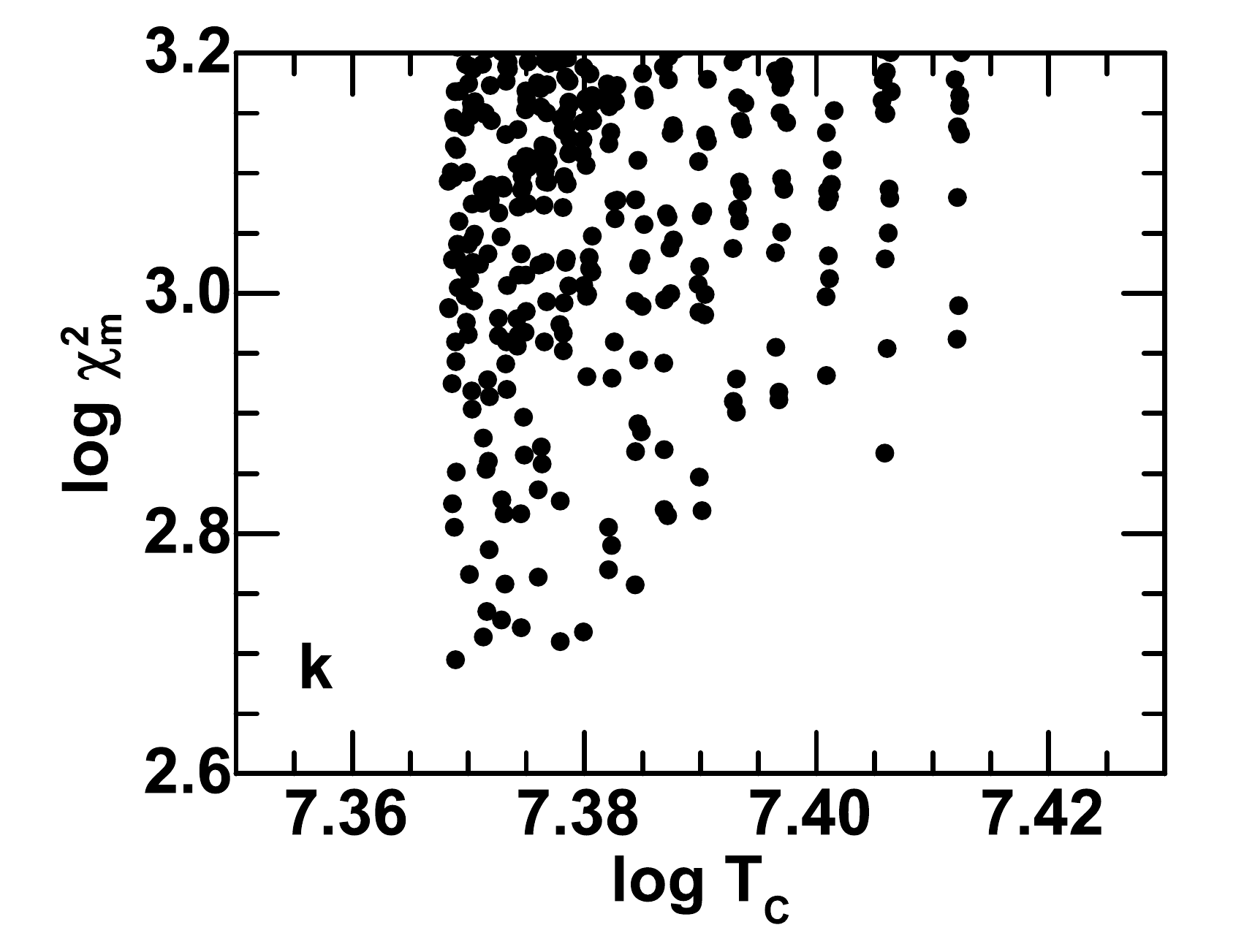}
	\includegraphics[width=0.66\columnwidth]{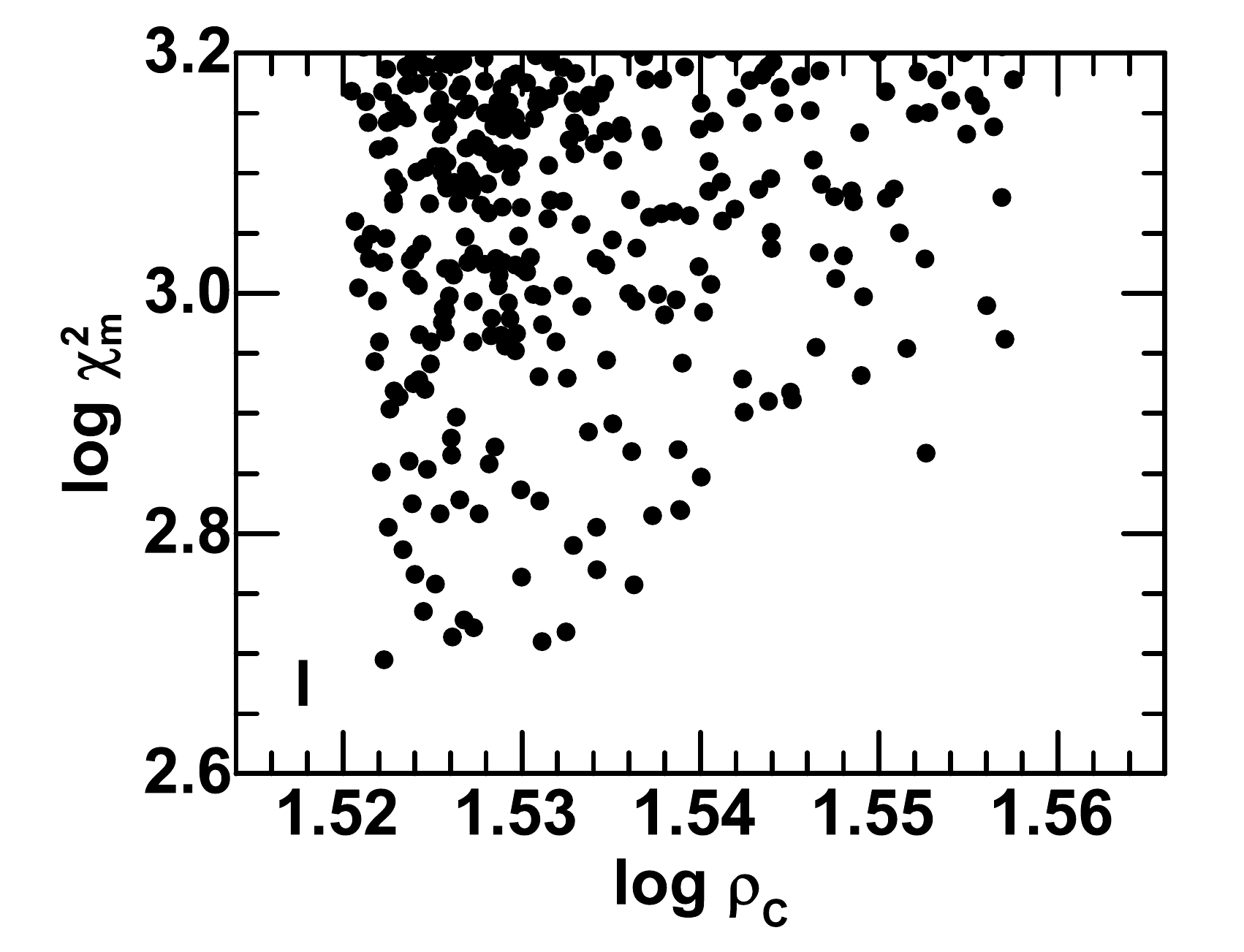}
    \caption{ Univariate investigation: $\log \chi^2_{\rm m}$ vs. single stellar parameter ($M$, $Z$, $\theta$, $C$) or stellar global variables (the stellar age, the effective temperature, the stellar radius, the center hydrogen abundance, the period spacing, the fraction of the convective core, the center temperature, and the center density). Only models in the fine grid with $\log \chi^2_{\rm m} < 3.2$ are shown. In panel d, the empty circles are the minimum of $\log \chi^2_{\rm m}$ at given $C$ obtained by interpolations of oscillation frequencies on other stellar parameters ($M$, $Z$, $\theta$ and the reduced stellar age $\tau$), see text. }
    \label{figpara1}
\end{figure*}

\begin{figure}
    \centering
	\includegraphics[width=\columnwidth]{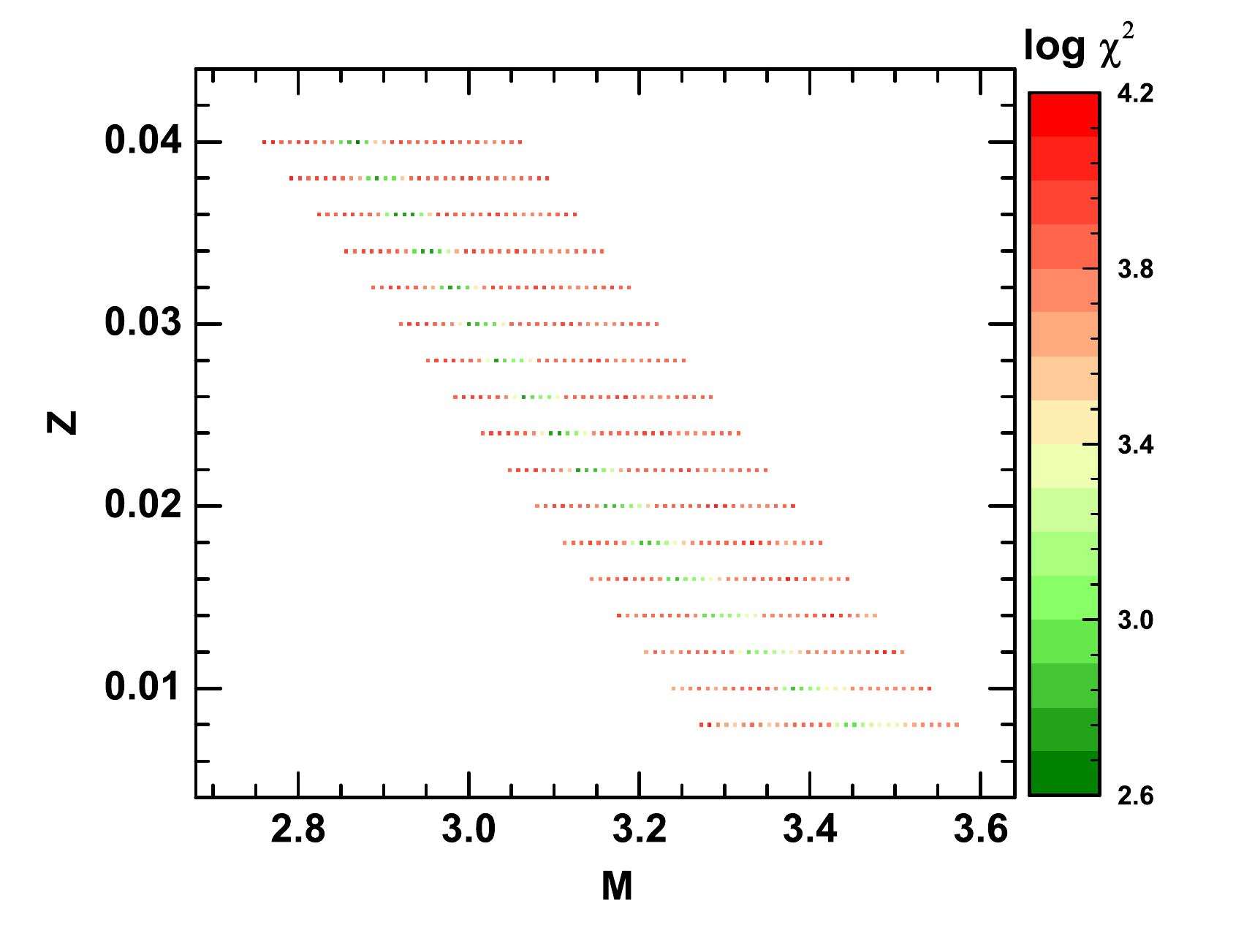}
	\includegraphics[width=\columnwidth]{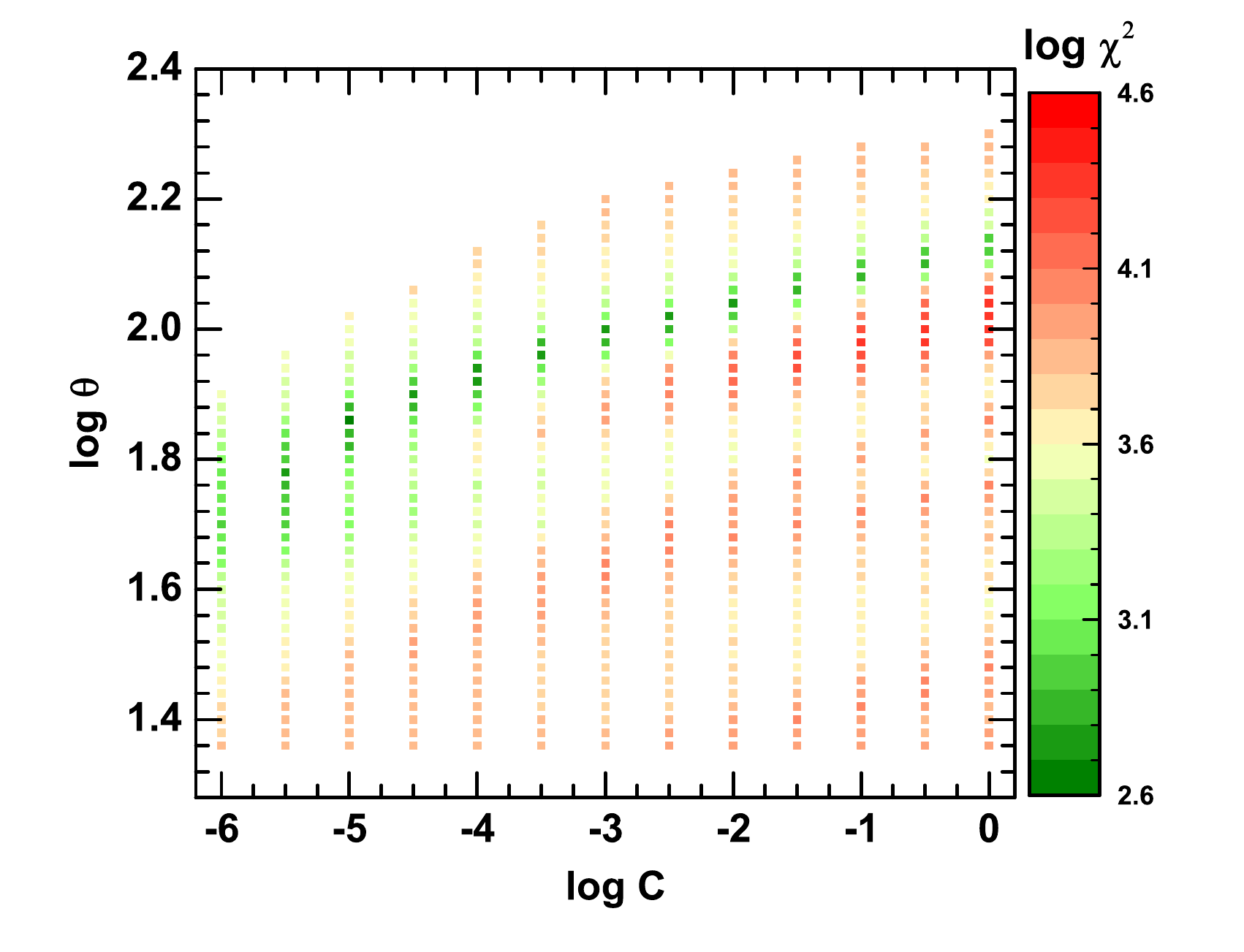}
    \caption{Similar with \ref{coarse}, but for models on the fine grid. }
    \label{figpara2}
\end{figure}

In Figure \ref{figpara1}a, it is shown that the suggested mass range for KIC 10526294 is 2.9-3.4 $M_{\odot}$, with the best value being approximately 3.0 $M_{\odot}$ based on the interpolation of oscillation frequencies. Within the considered range of $0.008 \leq Z \leq 0.040$, Figure \ref{figpara1}b demonstrates that the minimum value of $\log \chi^2_{\rm m}$ increases as $Z$ increases, with the minimum almost reaching the upper limit of the suggested range, 2.95, at $Z=0.010$. Therefore, the suggested range for metallicity is $Z \geq 0.01$. However, a higher $Z$ value larger than 0.02 is more likely since the minimum value of $\log \chi^2_{\rm m}$ is lower and exhibits minimal changes for $Z\geq0.02$. This preference for a higher $Z$ value is also supported by the non-adiabatic investigation on mode excitation \citep{Moravveji2015}.

In Figure \ref{figpara1}b, it is evident that the model with the lowest $\chi_m^2$ is obtained when $Z=0.04$. The reason for not using a higher upper limit for $Z$ is that the OPAL EOS tables implemented in the YNEV code are only valid for $0 \leq Z \leq 0.04$. Hence, models with $Z > 0.04$ are not included in this analysis to avoid extrapolations on EOS variables. Although the best model has $Z=0.04$, it can be observed from Figure \ref{figpara1}b that the minimum value of $\chi^2_{\rm m}$ changes little as $Z$ increases beyond 0.03. Consequently, setting a higher value for $Z$ is unlikely to significantly reduce $\chi^2_{\rm m}$. Moreover, the spectral results by \citet{Papics2014} indicate that a $Z$ value exceeding 0.04 is unlikely for KIC 10526294.

The two variables investigation on both $M$ and $Z$ is depicted in the top panel of Figure \ref{figpara2}. The color of each data point represents the minimum value of $\log \chi^2_{\rm m}$ for fixed $M$ and $Z$ and different $\log C$ and $\log \theta$. By applying the criterion $\chi^2_{\rm m}<2\chi^2_{\rm best}$, a tight constraint on $M$ and $Z$ is revealed as follows:
\begin{eqnarray} \label{fittingMZ}
\frac{M}{M_{\odot}}+16Z=3.50\pm0.03 \quad (\text{for } Z\geq0.016) \\ \nonumber
\frac{M}{M_{\odot}}+26Z=3.66\pm0.03 \quad (\text{for } Z<0.016).
\end{eqnarray}
This constraint aligns with the findings of \citet{Moravveji2015}. Negative correlations between mass and metallicity have been observed in asteroseismic modeling of other intermediate-mass stars as well, such as \citet{Ausseloos02} and \citet{Wu2020}. A plausible explanation, as suggested by \citet{Wu2020}, is that the combination of parameters $M$ and $Z$ should result in an appropriate convective core to match the observed period spacing pattern. In comparison to \citet{Moravveji2015}, the suggested mass in this study is slightly lower. This discrepancy is likely due to the wider range of $Z$ values employed in our fine grid. The constraint on $M$ and $Z$ indicates a lower stellar mass for higher $Z$. In \citet{Moravveji2015}, the maximum $Z$ value in their fine grid is 0.02, which is lower than ours. Their best model exhibits a stellar mass of approximately 3.25 $M_{\odot}$ and $Z=0.014$, consistent with the constraint in Equation (\ref{fittingMZ}) discovered in this work.

\begin{figure}
    \centering
	\includegraphics[width=\columnwidth]{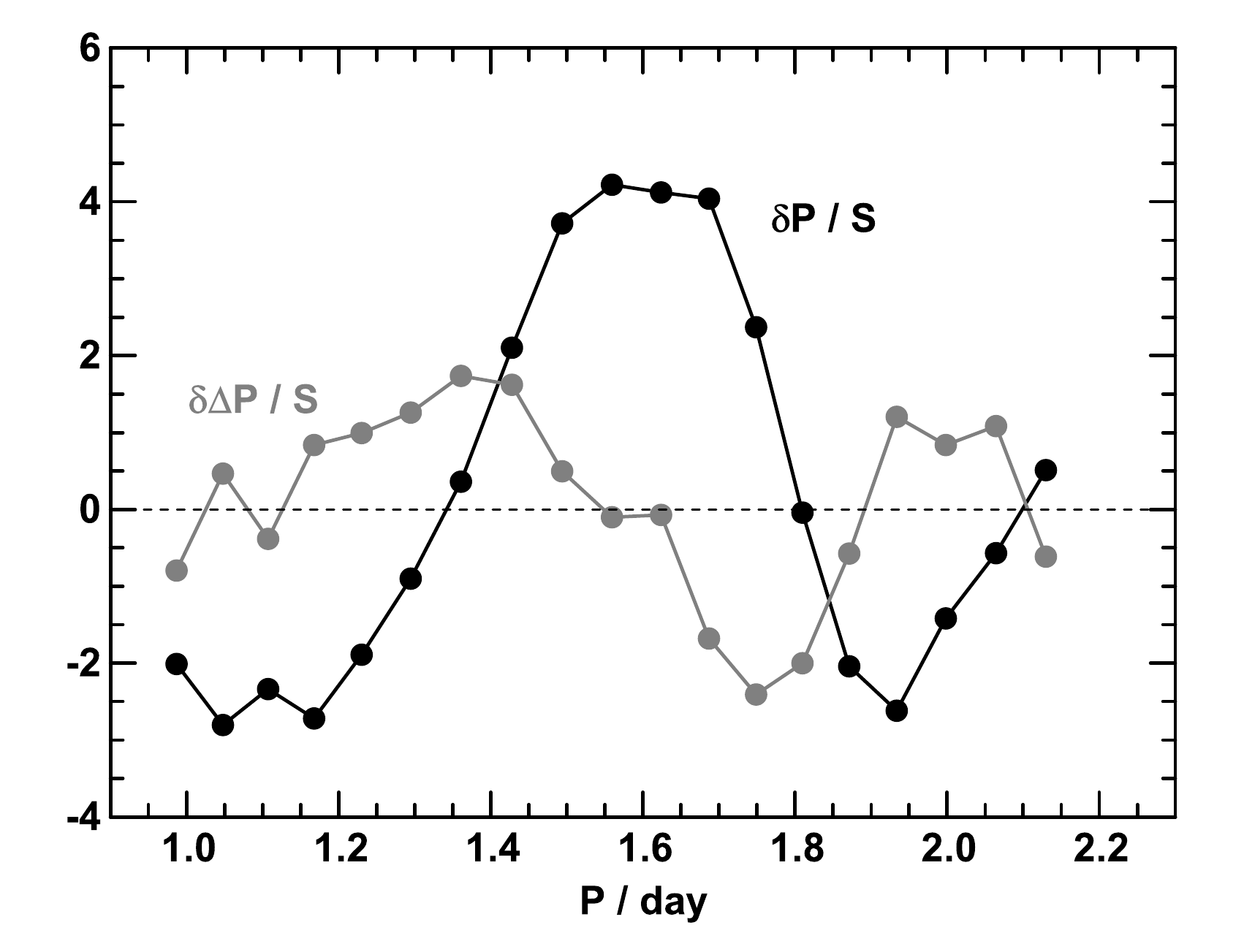}
    \caption{The differences of the oscillation periods (shown as the black symbols) and the period spacings (shown as the grey symbols) between interpolation and real calculations of stellar evolution and oscillation for the best model with the parameter $\log C=-4$, $\log \theta=1.932$, $M=2.983M_{\odot}$, $Z=0.03171$ and $\tau=0.1942$. The values of $\log \theta$, $M$, $Z$ and $\tau$ are obtained by using an iteration on them to search the overall minima of $\chi^2_{\rm m}$ based on the interpolation in the whole parameter space. }
    \label{figpara2vs}
\end{figure}

The overshoot parameter $C$ is of particular interest, and thus we conducted a detailed investigation of the best stellar model for a given $\log C$ by interpolating the oscillation frequencies. Interpolation of the frequencies from the calculated stellar models is necessary for the given values of $\log C$, $\log \theta$, $M$, $Z$, and reduced stellar age $\tau=t/t_{\rm ms}$ (where $t_{\rm ms}$ represents the stellar lifetime on the main sequence). The interpolation process involves two steps. In the first step, we interpolate the frequencies at an arbitrary $\tau_*$ on a stellar evolutionary track determined by the stellar parameters at a mesh point on the grid. Hermite interpolation with piecewise cubic polynomials is employed for this purpose. The derivatives of the frequencies with respect to $\tau$ are calculated using a 5-point Taylor expansion. The second step involves interpolating the frequencies using arbitrary $\log \theta_*$, $S_*$, $Z_*$, and $\tau_*$. For any given point $(\log \theta_*, S_* ,Z_*)$ within the considered range of the fine grid, we identify the elemental cell that contains it. We then interpolate the frequencies at $\tau_*$ on the eight stellar evolutionary tracks corresponding to the eight apexes of the cell, following the first step. Subsequently, trilinear interpolation is performed on $\log \theta$, $S$ and $Z$ to obtain the frequencies with the parameters $\log \theta_*$, $S_*$, $Z_*$, and reduced age $\tau_*$. This interpolation scheme ensures the continuity of frequencies throughout the entire space of stellar parameters and ages. By iterating based on the continuously interpolated frequencies, we can determine the minimum value of $\log \chi^2$ for a given $\log C$. To assess the accuracy of the interpolation scheme, we compare the differences in oscillation periods $\delta P$ and period spacings $\delta \Delta P$ between the results of interpolation and those obtained from actual calculations of stellar evolution and oscillation for the best model. The differences, displayed in Figure \ref{figpara2vs}, are at a level of approximately $1$ second, confirming the validity of the interpolation scheme.

The distribution of $\log \chi^2_{\rm m}$ for the mesh points on the grid with respect to $\log C$ is presented in Figure \ref{figpara1}d. The empty circles indicate the minimum values of $\log \chi^2_{\rm m}$ for given $\log C$ obtained using the interpolation method described earlier. It can be observed that the minima of $\log \chi^2_{\rm m}$ for all considered $\log C$ values are below 2.95, and the minimum value of $\log \chi^2_{\rm m}$ at $\log C = -6$ is close to 2.95. Thus, the suggested range for $\log C$ is $\log C \geq -6$. An important result from Figure \ref{figpara1}d is that the overall minimum of $\log \chi^2_{\rm m}$ does not occur at $\log C = 0$, but rather around $\log C = -4$, with minimal variation in the range $-5 \leq \log C \leq -3$. This finding suggests the existence of a rapid decrease in the diffusion coefficient in a thin layer near the convective boundary, which is a crucial aspect of the adopted overshoot mixing model. The diffusion mixing model proposed by \citet{Herwig2000}, corresponding to $\log C = 0$, does not yield the best result. Another noteworthy observation is that overshoot mixing must be taken into account because the minima of $\log \chi^2_{\rm m}$ rapidly increase as $\log C$ decreases when $\log C < -5$. As the overshoot mixing is proportional to $C$, very small values of $C$ indicate negligible overshoot.

However, it should be noted that $\log \chi^2_{\rm m}$ is not highly sensitive to changes in $\log C$. The reduction in $\chi^2_{\rm m}$ is only approximately $30\%$ when $\log C$ varies from 0 to -4. According to the adopted criterion, which sets $\chi^2_{\rm m}<2\chi^2_{\rm best}$ as the suggested range, $C=1$ still falls within the suggested range. Therefore, the results provide weak evidence for the rapid decrease of the diffusion coefficient near the convective boundary. Further investigations involving a larger sample size are required to confirm this observation.

The results of two variables investigation on $\log C$ and $\log \theta$ are presented in the bottom panel of Fig.\ref{figpara2}. Based on Fig.\ref{figpara1}c and the criterion $\chi^2_{\rm m}<2\chi^2_{\rm best}$, the suggested range for $\log \theta$ is $1.7 \geq \log \theta \geq 2.1$. In the classical diffusion mixing model, $\theta$ is linked to $f_{ov}$ through the relation $f_{ov}\theta=2$. Consequently, the suggested range for $f_{ov}$ is from 0.016 to 0.040. It is important to note that the broad range of $f_{ov}$ is due to the relatively loose constraint on $C$. However, if we focus on the case of $C=1$, which corresponds to the diffusion mixing model proposed by \citet{Herwig2000}, the result depicted in Fig.\ref{figpara2} indicates a tight constraint on $\log \theta$ as 2.12$\pm0.01$ for $\log \chi^2_{\rm m}<2.95$, corresponding to $f_{\rm ov}=0.016$. This finding is consistent with the results reported by \citet{Moravveji2015} and also aligns with the recommendation made by \citet{Herwig2000}.

The distributions of $\log \chi^2_{\rm m}$ are shown in Fig.\ref{figpara1}e-i for various stellar global variables, including the stellar age $t$, effective temperature $T_{\rm eff}$, radius $R$, center hydrogen abundance $X_C$, period spacing $\Delta P$, mass fraction of the convective core mass $M_C/M$, center temperature $T_C$, and center density $\rho_C$. The suggested range and the best value for these variables are listed in Table \ref{tablepararange}. Additionally, the suggested range for the surface gravitation acceleration is determined to be $4.15 \leq \log g \leq 4.25$ in c.g.s. units. The results obtained for $T_{\rm eff}$ and $\log g$ are consistent with the spectral analysis conducted by \citet{Papics2014}. It is worth noting that the period spacing is the global variable that exhibits the tightest constraint. This result is expected as high-order g-mode oscillations tend to exhibit nearly uniform spacing in periods.

\begin{table}
\centering
\caption{ The suggested range and the best values of the stellar parameters and global variables, in c.g.s unit. }\label{tablepararange}
\begin{tabular}{lcc}
\hline\noalign{\smallskip}
       global variable                & suggested range                & best     \\
\hline
          $M/M_{\odot}$                         & 2.9  $\sim$  3.4     &  3.0    \\
          $Z$                         & 0.01 $\sim$ 0.04     &  0.04    \\
          $\log \theta$               & 1.7 $\sim$ 2.1       &  1.9     \\
          $\log C$                    & $\geq$-6             &  -4      \\
          $t$ / Gyr                   & 0.05 $\sim$ 0.09     &  0.08    \\
        $\log T_{\rm eff}$            & 4.02 $\sim$ 4.16     &  4.04    \\
        $R / R_{\odot}$               & 2.12 $\sim$ 2.38     &  2.36    \\
        $X_C$                         & 0.56 $\sim$ 0.65     &  0.57    \\
        $\Delta P / s$                & 5500 $\sim$ 5600     &  5570    \\
        $M_C / M$                     & 0.18 $\sim$ 0.22     &  0.19    \\
        $\log T_C$                    & 7.37 $\sim$ 7.41     &  7.37    \\
        $\log \rho_C$                 & 1.52 $\sim$ 1.56     &  1.52    \\
\hline
\end{tabular}
\end{table}

\subsection{The differences between the models and the observations} \label{SecResult4}

\begin{figure}
    \centering
	\includegraphics[width=\columnwidth]{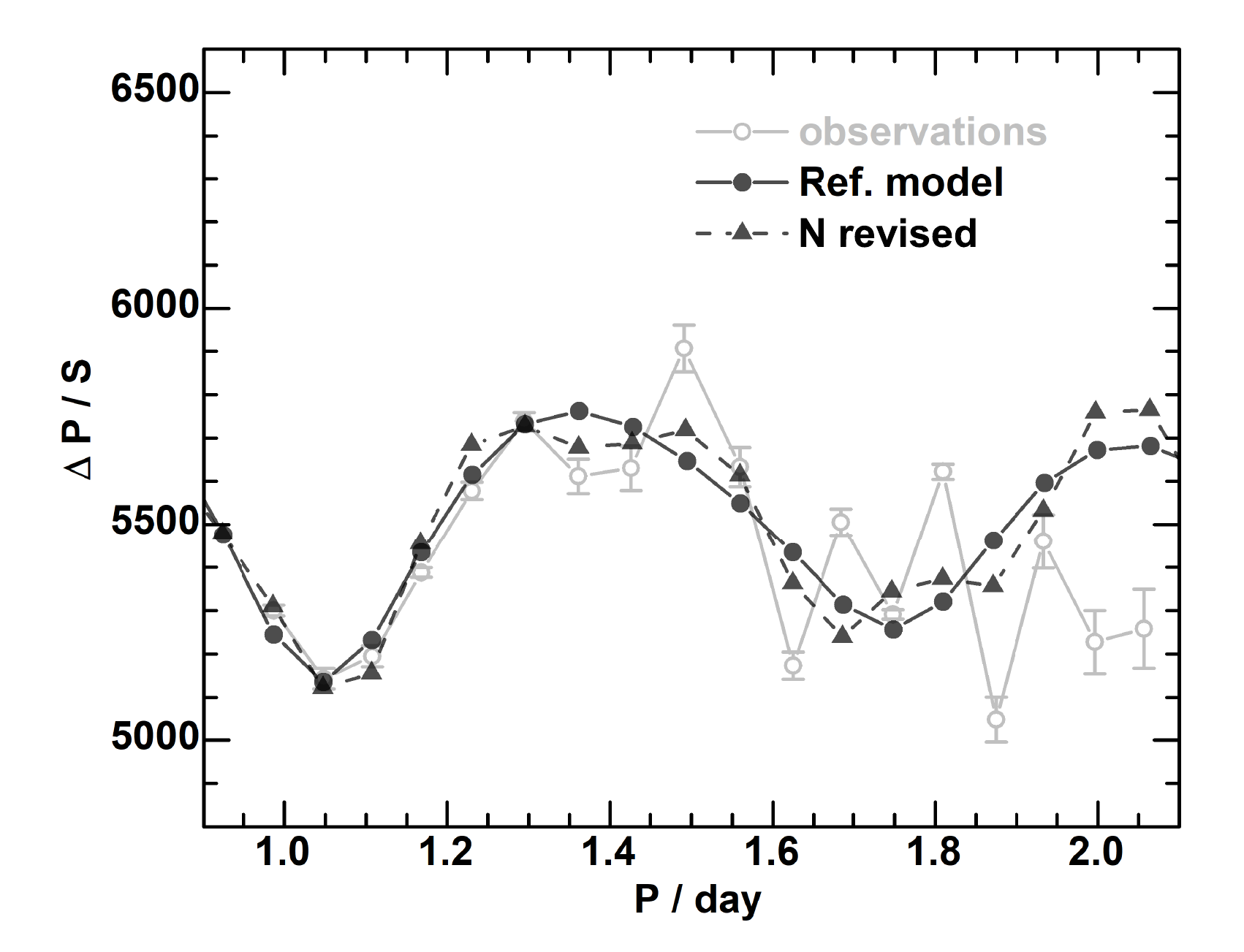}
    \caption{The period spacings of the observations with the error bas multiplied by three are shown by the grey symbols. The period spacings of the reference model are shown by the black dots linked with the solid line. The period spacings of the model with revised $N$ profile are shown by the black triangles linked with the dashed line. }
    \label{figdperr}
\end{figure}

\begin{figure}
    \centering
	\includegraphics[width=\columnwidth]{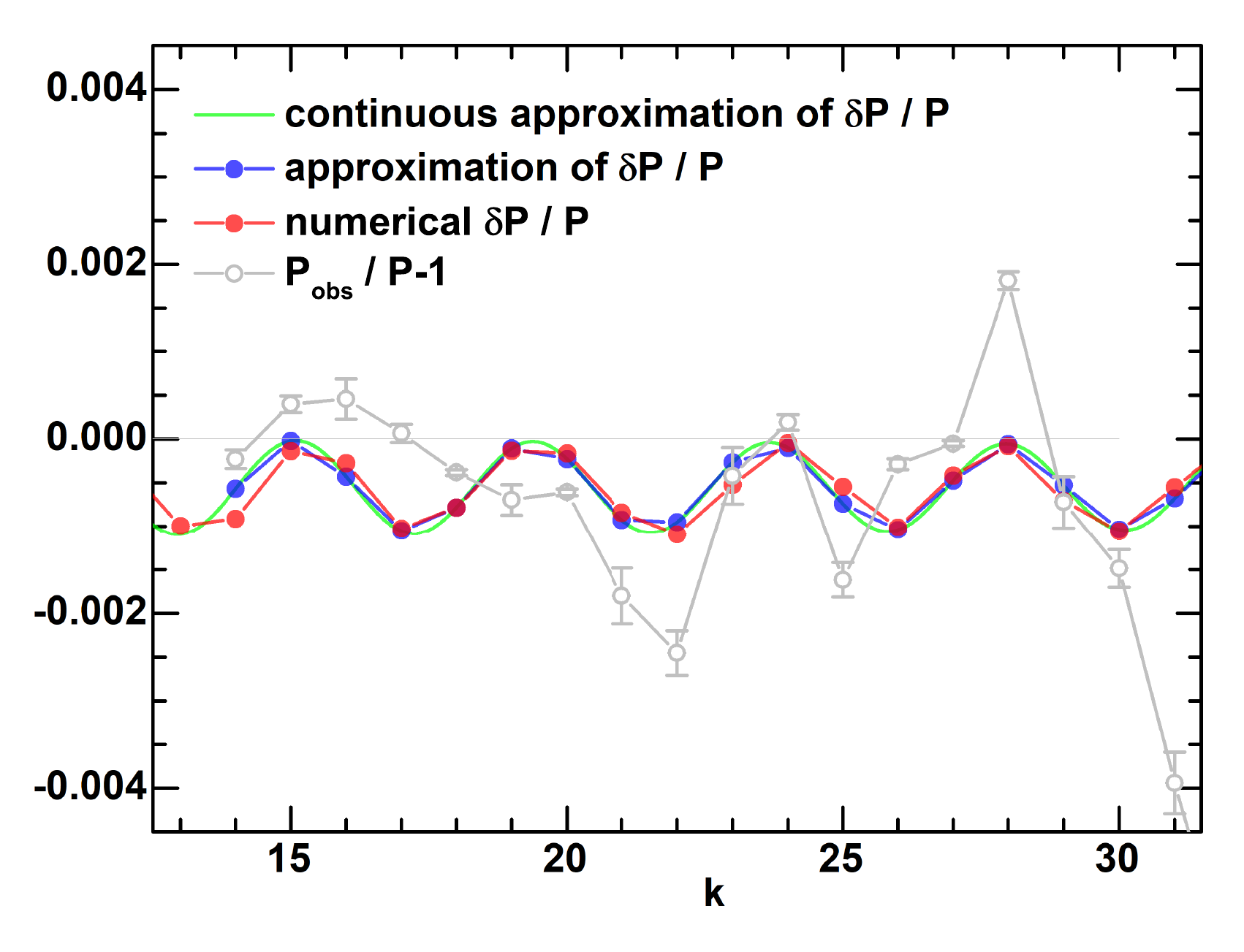}
    \caption{ Relative differences of oscillation periods among the reference model, the model with revised $N$ profile and the observations, also the analytical approximation of variation of oscillation periods. }
    \label{figperiodfit}
\end{figure}

The minimum value of $\chi^2$ is still significantly larger than unity, indicating a substantial residual in the oscillation periods. In Fig.\ref{figdperr}, the period spacings of the lowest-$\chi^2$ model are represented by black dots, while the observed period spacings are denoted by grey empty circles. A comparison between the two reveals that the model's period spacing exhibits a generally smooth pattern, whereas the observations display irregular oscillatory behavior. The relative differences between the observed oscillation periods and those of the lowest-$\chi^2$ model are depicted by the grey symbols in Fig.\ref{figperiodfit}. This plot demonstrates that the irregular oscillatory behavior of the period spacings arises from similar variations in the oscillation periods themselves. To gain insight into the observed pattern of oscillation periods, it is useful to investigate potential adjustments to the stellar models by analyzing the residuals between a reference stellar model and the observed periods.

The Cowling approximation simplifies the stellar adiabatic oscillation equations to second-order equations, allowing for analytical investigations using the JWKB approximation for high-order modes \citep[see, e.g.,][]{Gough1993, Montgomery2003}. By combining this approximation with the asymptotic expression of the $g$-mode period \citep{Tassoul1980}, given by ${P_k} \approx 2{\pi ^2}{\Pi _0}(k + {k_0})/L$ (where $k$ represents the mode node and $L=\sqrt{l(l+1)}$), and employing the variational method, we can derive the variations in the oscillation period resulting from a perturbation of the Brunt-V\"{a}is\"{a}l\"{a} frequency $N$ as
\begin{eqnarray} \label{zqsdPP}
\frac{{\delta {P_k}}}{{{P_k}}} \approx  - \int\limits_0^1 {z(u)\{ 1 + \sin [2\pi (k + {k_0})u]\} du},
\end{eqnarray}%
where
\begin{eqnarray} \label{zqsdPP1}
z = \frac{{\delta N}}{N},
\end{eqnarray}%
\begin{eqnarray} \label{zqsdPP2}
x = \frac{{r}}{R},
\end{eqnarray}%
\begin{eqnarray} \label{zqsdPP3}
{\Pi _x}^{ - 1} &=& \int\limits_{0}^x {\frac{{\left| N \right|}}{x}dx},
\end{eqnarray}%
\begin{eqnarray} \label{zqsdPP4}
\Pi _0 = \Pi _x(x=1),
\end{eqnarray}%
and
\begin{eqnarray} \label{zqsdPP5}
u = \frac{{{\Pi _x}^{ - 1}}}{{{\Pi _0}^{ - 1}}}.
\end{eqnarray}%
Equation (\ref{zqsdPP}) differs slightly from the result presented in \citet{Miglio2008}. The detailed derivation of Eq.~(\ref{zqsdPP}) and its numerical validation are provided in the Appendix. It has been shown that this equation accurately reproduces both the qualitative and quantitative aspects of the numerical results concerning the variation of the oscillation period.

Considering a perturbation of $N$ in a single region with $u_1 \leq u \leq u_2$, let $z(u_1)=z(u_2)=0$ for continuity and expand $z$ to Fourier sine series:
\begin{eqnarray} \label{zexp}
z(u) = \left\{ {\begin{array}{*{20}{c}}
   {0, u < {u_1} }  \\
   {\sum\limits_{n = 1}^\infty  {{A_n}\sin (\frac{{u - {u_1}}}{{{u_2} - {u_1}}}n\pi )} , {u_1} \le u \le {u_2}}  \\
   {0, u > {u_2} }  \\
\end{array}} \right.,
\end{eqnarray}%
where $A$ is the coefficients of the sine series. The variation of oscillation period based on Eq.(\ref{zqsdPP}) is
\begin{eqnarray} \label{zqsdPP5}
&&\frac{{\delta {P_k}}}{{{P_k}}} \approx  - \frac{{({u_2} - {u_1})}}{\pi }\sum\limits_{n = 1}^\infty  {\frac{{{A_n}}}{n}} \{ [1 - {( - 1)^n}] + \\ \nonumber
&& \frac{{\sin [2\pi (k + {k_0}){u_1}] - {{( - 1)}^n}\sin [2\pi (k + {k_0}){u_2}]}}{{1 - {{[\frac{{2(k + {k_0})({u_2} - {u_1})}}{n}]}^2}}}\}.
\end{eqnarray}%
When perturbation of $N$ exists in many detached regions, the total $\delta P_k / P_k$ is the sum of $\delta P_k / P_k$ in each regions.

By utilizing the above formula, it becomes possible to investigate the potential modifications to the $N$ profile that may correspond to the characteristics of the oscillatory component in the observed period spacing. Using a reference model to calculate $P_k$, the objective is to find the values of $u_1$, $u_2$, and $A_n$ for all regions that minimize the following function:
\begin{eqnarray} \label{revQ}
Q = \sum\limits_k {{{\left(\frac{{\delta {P_k}}}{{{P_k}}} - \frac{{{P_{k,\rm{obs}}} - {P_k}}}{{{P_k}}}\right)}^2}}.
\end{eqnarray}
In this equation, $\delta P_{{\rm{obs}},k}$ is not used as the denominator since the uncertainty arising from Eq. (\ref{zqsdPP}) based on the asymptotic relation is not comparable to the very small observational uncertainties. Considering $\delta P_{{\rm{obs}},k}$ as the denominator would introduce significant random variations to $Q$, rendering it unhelpful for analysis. Despite not incorporating $\delta P_{{\rm{obs}},k}$, the above equation is sufficient to capture the observational characteristics of the period spacing through variations in the reference model. In this problem, all $A_n$ can be determined using the linear least square method, thus only $u_1$ and $u_2$ for all regions need to be determined.

\begin{figure}
    \centering
	\includegraphics[width=\columnwidth]{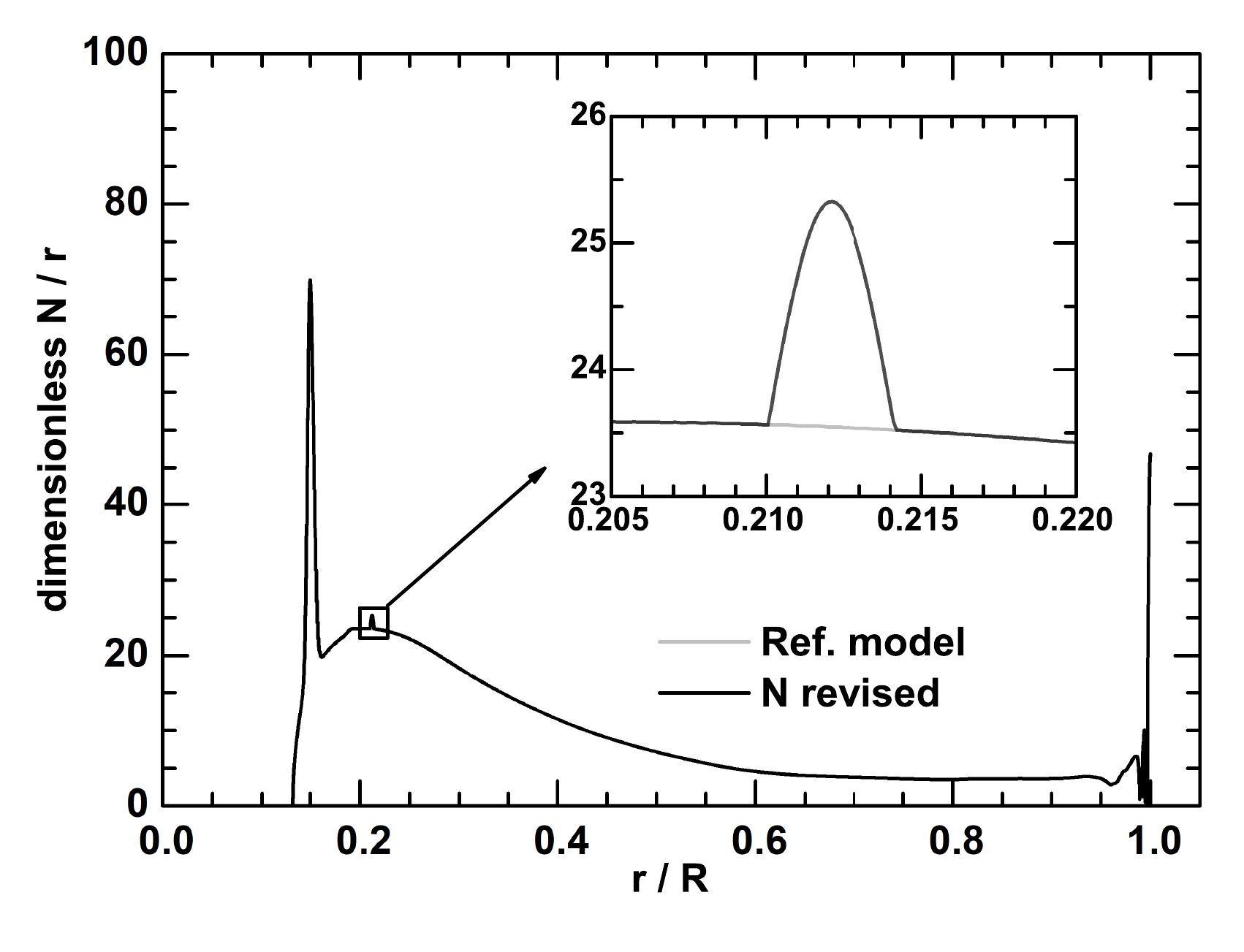}
    \caption{Revised profile of $N$ constructed by perturbation $z$ in two regions determined by Eq.(\ref{zexp}) with the coefficients in text. }\label{figdNNrev}
\end{figure}

In our case, we have selected the stellar model with the following parameters as the reference model: $\log C=-4$, $\log \theta=1.932$, $M=2.983M_{\odot}$, $Z=0.03171$, and $\tau=0.1942$. For this model, $k_0$ is fixed at 1.0. Alternatively, $(k+k_0)$ in Eq. (\ref{zqsdPP}) and Eq. (\ref{zqsdPP5}) can be replaced with $P_k/\Delta P$, where $\Delta P = 2{\pi ^2}{\Pi _0}/L$ represents the asymptotic period spacing. The differences between the two approaches are negligible. Perturbations on $N$ within a single region are assumed, and only the lowest-order term in the Fourier sine series is taken into account. By scanning through different values of $u_1$ and $u_2$, it was found that the profile of $\delta N / N$ that minimizes $Q$ is determined by $A_1=0.0755$ within the range $0.2273 \leq u \leq 0.2388$. The revised profile of $N$ is illustrated in Fig.\ref{figdNNrev}. It should be noted that the three observational modes with the highest periods are not taken into account, as their inclusion would result in Fourier sine series coefficients that are too large and violate the assumption of perturbation.

The results of the relative variations in the oscillation periods with the $\delta N / N$ profile are depicted in Fig.\ref{figperiodfit}. In this context, $P$ represents the oscillation period of the reference model, $P_{\rm rev}$ signifies the oscillation period of the reference model with the revised $N$ profile, and $\delta P = P_{\rm rev} - P$. The revised $N$ profile leads to a reduction in the overall residual of the oscillation periods, i.e., $\sum\limits_k {\left| {{P_{{\rm rev},k}} - {P_{{\rm obs},k}}} \right|} = 3208$ s, which is lower than $\sum\limits_k {\left| {{P_k} - {P_{{\rm obs},k}}} \right|} = 3707$ s. However, the revised $N$ profile yields a higher $\chi^2 = 736$ compared to $\chi^2 = 692$ for the reference model. The analytical approximation Eq.(\ref{zqsdPP}) is not accurate enough to comparable with the very small observational uncertainties ($\Delta P_{\rm{obs}}/P_{\rm{obs}} $ is less than about $ 10^{-4}$), so that $\chi^2$ is out of control. The differences in oscillation periods between observations and the reference model, $P_{\rm obs}/P - 1$, are illustrated by the grey symbols, while the error bars are multiplied by three to enhance their visibility. The analytical $\delta P/P$ based on Eq. (\ref{zqsdPP5}) is represented by the blue symbols, and the numerical $\delta P/P$ calculated using the oscillation code is shown as the red symbols. The numerical $\delta P/P$ generally reproduces the analytical results, validating the reasonability of the approximation given by Eq. (\ref{zqsdPP}). However, the numerical $\delta P/P$ does not perfectly replicate $P_{\rm obs}/P - 1$ since the former exhibits regular oscillatory behavior while the latter is irregular. This indicates that the irregular oscillatory behavior observed in the period spacings cannot be attributed to a simple sine function of $\delta N/N$ in a single region. Nevertheless, the numerical $\delta P/P$ qualitatively reveals the oscillatory nature of the observational period spacings, as the positions of the peaks and valleys in the numerical $\delta P/P$ align with those in $P_{\rm obs}/P - 1$. Similar results can be observed in the period spacings of the model with the revised $N$ profile, illustrated in Fig.\ref{figdperr} as black triangles connected by a dashed line. It can be found that the model with the revised $N$ profile exhibits an oscillatory component, and its period spacings are closer to the observations than those of the reference model within the range of $1.3<P/{\rm day}<1.9$. It is evident from Eq. (\ref{zqsdPP}) that the period (measured by the node number $k$ as shown in Fig.\ref{figperiodfit}) of the oscillatory behavior is sensitive to the location of the revised $N$ profile. Therefore, the oscillatory behavior observed in the period spacings suggests the existence of an additional abundance gradient near $u=0.23$ or $r=0.21R$.

The above investigation suggests that the stellar models lack additional physical processes that could cause variations in $N$ near $r=0.21R$. Some potential mechanisms directly associated with this matter include mixing and inhomogeneous accretion during the PMS phase. Exploring these possible mechanisms would require extensive calculations of numerous stellar models. However, investigating these aspects is beyond the scope of this paper, and we intend to address them in future research.

\section{Discussions} \label{SecDis}

\subsection{Implications on the convective overshoot}

As shown by Zhang et al. (2013), the diffusion coefficient for convective mixing is simplified as $D_{\rm CZ} = C_1 k \tau$ within the convection zone and $D_{\rm OV} = C_2 k / (\tau N^2)$ within the overshoot region. Here, $k$ represents the turbulent kinetic energy, $\tau$ is the timescale of turbulent dissipation, and $C_1$ and $C_2$ are parameters. The distinct expressions for $D$ in the convection zone and overshoot region arise from the different characteristic lengths of turbulent motion, specifically $L_{\rm CZ} \propto \sqrt{k} \tau$ and $L_{\rm OV} \propto \sqrt{k} / N$. The parameter $C$ in the adopted overshoot mixing model, Equation (\ref{OVMZCDL2022}), measures the ratio of $D_{\rm OV}$ to $D_{\rm CZ}$ for the same $k$, i.e., $C = C_2 / (C_1 \tau^2 N^2)$. The turbulent dissipation timescale $\tau$ multiplied by the typical value of $N$ in the overshoot region is usually much larger than unity. For the best model of KIC 10526294, $L_{\rm CZ}$ is constrained by the size of the convective core, yielding $L_{\rm CZ} \approx 2 \times 10^{10}$ cm. The estimated typical turbulent velocity near the convective boundary, using the MLT, is approximately $\sqrt{k} \approx 2 \times 10^3$ cm/s, resulting in $\tau = L_{\rm CZ} / \sqrt{k} \approx 10^7$ s. Additionally, the typical value of $N^2$ just outside the convective core is approximately $N^2 \approx 10^{-7}$ / s$^2$, giving ${\tau^2}{N^2} \approx 10^7 \gg 1$. Consequently, the turbulent convective mixing model (Zhang et al., 2013) predicts a rapid decrease in characteristic length and diffusion coefficient near the convective boundary, corresponding to $C \ll 1$. The asteroseismic analysis of KIC 10526294 reveals a best-fit value of $C = 10^{-4}$, while $C = 1$ is also within an acceptable range, providing weak evidence to support the model prediction.

However, changing $\log C$ from $0$ to $-4$ only results in a modest reduction of approximately $30\%$ in $\chi^2$. This lack of significant improvement can be attributed to two main reasons. Firstly, the strength of the overshoot mixing is primarily determined by the exponential decay factor $\theta$, making it challenging to sensitively investigate the effects of varying $C$. To obtain a more conclusive understanding of convective overshoot, further investigations encompassing a broader range of stellar types and samples are necessary to confirm the value of $C$. Secondly, there are additional physical processes (such as convective heat flux in the overshoot region, non-adiabatic and non-linear effects in stellar oscillations, and abnormalities in the $N$ profile) that are not accounted for in the stellar models but correspond to observed oscillatory period spacing patterns. As a result, a relatively high lower limit for $\chi^2$ exists, constraining the potential improvement in $\chi^2$ solely attributed to overshoot mixing.

If the rapid decrease of the convective characteristic length for the mixing of abundance and entropy near the convective boundary is valid, it raises a question regarding the discrepancy between the convective characteristic lengths for thermal-dynamic quantities and dynamic quantities in the overshoot region. The P\'{e}clet number is typically defined as the ratio of the convective diffusion coefficient of kinetic energy to the radiative diffusion coefficient. Therefore, $Pe\gg1$, which corresponds to adiabatic stratification, is only achieved if the convective diffusion coefficient of kinetic energy is equal to that of entropy. It is important to note that in statistical turbulent convection models, the convective transport of a thermal-dynamic quantity (such as abundance or entropy) is described by a second-order correlation, whereas the convective transport of a dynamic quantity (such as turbulent kinetic energy or turbulent heat flux) is described by a third-order correlation. When these correlations are modeled using the down-gradient approximation, it is not necessary for the resulting diffusivity to be the same. In current statistical turbulent convection models, the second-order correlations for the convective transport of thermal-dynamic quantities ($\overline{u'_rT'}$ and $\overline{u'_rX'}$) have been found to be leaded by a gradient term \citep[see, e.g.,][]{Zhang2012b,Zhang2013}. However, numerical simulations of convective overshoot have demonstrated that the down-gradient approximation is not always appropriate for the third-order correlation of convective transport \citep[e.g.,][]{Chan1996,Kupka2007,Tian2009}. This comparison also suggests a potential difference in the characteristic lengths between thermal-dynamic quantities and dynamic quantities in the overshoot region.

\subsection{Effects of varied initial abundance and extra global mixing}

In this study, we made the assumption $Y=0.2485+2Z$ for initial abundances to avoid excessively long calculation times for the stellar models. However, this assumption restricts the freedom of the stellar models. This is similar to the approach taken by \citet{Moravveji2015}, where the initial hydrogen abundance was fixed. Limiting this freedom may lead to an overestimation of the correlations between stellar parameters, such as $M$ vs. $Z$ and $\log C$ vs. $\log \theta$ shown in Fig.\ref{figpara2}. Since our work and \citet{Moravveji2015} employed different constraints on the initial abundance, the effects of removing these constraints can be estimated by considering the uncertainties associated with the correlations between pairs of parameters. Taking the example of $M$ vs. $Z$, Fig.\ref{figpara2} illustrates an uncertainty of approximately 0.03 solar masses in stellar mass for a given $Z$, whereas Fig.5 in \citet{Moravveji2015} shows an uncertainty of stellar mass of less than 0.02 solar masses for a given $Z$ (using the same criteria of $\chi^2< 2\chi^2_{\rm best}$ for a given $Z$ as the recommended parameter range). If we remove the constraint on the initial abundance, the correlations between stellar parameters depicted in Fig.\ref{figpara2} would be weaker, resulting in a larger dispersion. For instance, the dispersion between $M$ and $Z$ would increase to approximately 0.04 ($\approx\sqrt{0.03^2+0.02^2}$).

The extra global mixing is held constant in this study. We did not explore varying values of ${\rm log} D_{\rm ext}$ also due to the computational demands associated with a large number of stellar models. It has been found that the optimal value of the extra global mixing is primarily determined by the period spacing pattern in the high-frequency range, which is related to the $\mu$-gradient. On the other hand, overshoot primarily affects the period spacing pattern in the low-frequency range \citep{Moravveji2015, Moravveji2015b}. Therefore, the extra global mixing and overshoot mixing are largely independent of each other. Consequently, the optimal value of the diffusion coefficient for the extra global mixing should remain unchanged when different overshoot models are employed. In this case, adopting the best value of ${\rm log} D_{\rm ext}$ as determined by \citet{Moravveji2015} should not introduce a systematic bias in the investigation of overshoot mixing.

\section{Conclusions} \label{SecConclusion}

In this study, we examined convective core overshoot mixing using asteroseismology of the SPB star KIC 10526294. The selected overshoot mixing model is based on the work of \citet{ZCDL2022}. The power-law diffusion coefficient is derived from numerical simulations and turbulent convection models, which provide a power-law description of the turbulent velocity (as discussed in Section \ref{SecIntro}). The parameter $C$ is determined based on the theoretical model of convective overshoot mixing presented by \citet{Zhang2013}, which predicts a different characteristic length in the overshoot region compared to the convection zone. The chosen overshoot mixing model aligns with the widely used model proposed by \citet{Herwig2000} when the parameter $C$ is set to 1.

The oscillation periods of the stellar models for KIC 10526294, generated by varying stellar parameters, are compared with the observed periods. Initially, a coarse grid with large parameter steps is employed to explore the preliminary parameter space. Subsequently, a finer grid with smaller parameter steps within the preliminary parameter space is used for a detailed analysis to constrain the overshoot parameters and stellar parameters. To address the issue of resolution in stellar parameters and age \citep[e.g.,][]{Wu2016}, an interpolation technique is applied to the oscillation frequencies, considering all parameters and stellar age. Hermite cubic polynomial interpolation is utilized for stellar age along a stellar evolutionary track, while linear interpolation is employed for stellar parameters and overshoot parameters. The estimated error in the oscillation period resulting from the interpolation scheme is within a few seconds.

Based on the comparison between the model oscillation periods and the observations, the stellar parameters and overshoot parameters are constrained. The suggested parameter ranges are listed in Table \ref{tablepararange}. While there are slight differences in the stellar parameters compared to the results of \citet{Moravveji2015}, our models tend to prefer a lower stellar mass. This discrepancy may be attributed to the broader range considered for the stellar parameters $Z$. The minimum value of $\chi^2$ in our results is approximately 442, which is lower than that obtained by \citet{Moravveji2015}. The improvement can be attributed to two main factors: the incorporation of the parameter $C$ variation and the interpolation of the oscillation frequencies using the parameters and stellar age.

Although the recommended range for the parameter $C$ still includes unity, the most intriguing finding is that the best value of $C$ is $10^{-4}$, significantly smaller than unity. This result could be viewed as weak evidence supporting the prediction of the turbulent convective mixing model proposed by \citet{Zhang2013}, which suggests a rapid decrease in the diffusion coefficient of convective mixing near the convective boundary.

Furthermore, an analysis of the residuals in the oscillation periods between a reference model and the observations is conducted. A formula describing the variation of the oscillation periods due to perturbations in the Brunt-V\"{a}is\"{a}l\"{a} frequency is derived and quantitatively validated. The location of the perturbation in the $N$ profile can be determined by comparing the model and observational periods, utilizing Fourier expansion for the perturbation and the linear least squares method for the expansion coefficients. It is found that the oscillatory behavior in the residuals of the period spacing may be related to anomalies in the $N$ profile. This method is expected to be applicable for analyzing the residuals of oscillation periods in general, comparing models with observations.

\acknowledgments

Many thanks to the anonymous referee for providing valuable comments which improved the original version.
Fruitful discussions with Prof. J{\o}rgen Christensen-Dalsgaard are highly appreciated. Funding for Yunnan Observatories is co-sponsored by the National Key R\&D Program of China (Grant No. 2021YFA1600400 / 2021YFA1600402), the Strategic Priority Research Program of the Chinese Academy of Sciences (grant No. XDB 41000000), the National Natural Science Foundation of China (grant No. 11773064, 11873084, 12133011, 12273104 \& 12288102), the Foundation of the Chinese Academy of Sciences (Light of West China Program and Youth Innovation Promotion Association), and the Yunnan Ten Thousand Talents Plan Young \& Elite Talents Project, and International Centre of Supernovae, Yunnan Key Laboratory (No. 202302AN360001).

\appendix

\section{Validation the equation of the variation of oscillation period for a perturbation of the Brunt-V\"{a}is\"{a}l\"{a} frequency }

Equation (\ref{zqsdPP}) is obtained using the variational method for a perturbation on the linear operator of the stellar oscillation. With the Cowling approximation and the JWKB approximation, the variational method shows that the variation of the $g-$mode oscillation frequency is given by \citep[see,e.g.,][]{Gough1993,Montgomery2003}
\begin{eqnarray} \label{Adom}
\frac{{\delta {\omega _k}}}{{{\omega _k}}} \approx  \frac{2}{{\int\limits_0^R {\left| N \right|dr'/r'} }}\int\limits_0^R {\frac{{\delta N}}{N}\frac{N}{r}dr{{\sin }^2}(\frac{{{P_k}}}{{\Delta P}}\pi \frac{{\int\limits_0^r {\left| N \right|dr'/r'} }}{{\int\limits_0^R {\left| N \right|dr'/r'} }} + \frac{\pi }{4}} ).
\end{eqnarray}
Here we use $P_k/\Delta P$ to replace $k-1/2$ in the original formula in \citet{Montgomery2003} in order to better fit the model oscillation periods to the asymptotical relation, i.e., ${P_k} \approx 2{\pi ^2}{\Pi _0}(k + {k_0})/L$ \citep{Tassoul1980}, where $k_0$ is generally not $-1/2$. Therefore the variation of the oscillation period is
\begin{eqnarray} \label{Adpp}
 \frac{{\delta {P_k}}}{{{P_k}}} && \approx  - \frac{{\delta {\omega _k}}}{{{\omega _k}}} \approx  - 2{\Pi _0}\int\limits_0^1 {\frac{{\delta N}}{N}\frac{N}{x}dx{{\sin }^2}(\pi \frac{{{P_k}}}{{\Delta P}}\frac{{{\Pi _x}^{ - 1}}}{{{\Pi _0}^{ - 1}}} + \frac{\pi }{4}} ) \\ \nonumber
  && =  - 2{\Pi _0}\int\limits_0^1 {\frac{{\delta N}}{N}\frac{N}{x}dx\frac{1}{2}[1 - \cos (2\pi \frac{{{P_k}}}{{\Delta P}}\frac{{{\Pi _x}^{ - 1}}}{{{\Pi _0}^{ - 1}}} + \frac{\pi }{2}} )] \\ \nonumber
  && =  - {\Pi _0}\int\limits_0^{{\Pi _0}^{ - 1}} {\frac{{\delta N}}{N}d{\Pi _r}^{ - 1}[1 + \sin (2\pi \frac{{{P_k}}}{{\Delta P}}\frac{{{\Pi _x}^{ - 1}}}{{{\Pi _0}^{ - 1}}}} )] \\ \nonumber
  && =  - \int\limits_0^1 {z(u)[1 + \sin (2\pi \frac{{{P_k}}}{{\Delta P}}u} )]du \\ \nonumber
  && \approx  - \int\limits_0^1 {z(u)\{ 1 + \sin [2\pi (k + {k_0})u} ]\} du.
\end{eqnarray}

We validate Eq. (\ref{Adpp}) in three cases: (i) a constant $z = \delta N/N$ within a region, (ii) a linear function of $z$ within a region, and (iii) a sine function of $z$ within a region. The reference model used for validation has a stellar mass of 2.983 solar masses, $Z=0.0317$, $\log \theta=1.932$, $\log C=-4$, and $\tau=0.1942$. However, it should be noted that Eq. (\ref{Adpp}) is independent of specific stellar models. To quantitatively validate Eq. (\ref{Adpp}), we calculate the $l=1$ oscillation periods $P_k$ for models with different perturbations of $z$ applied to the reference model. We then compare the quantity $P_k/P_{k,0}-1$ with Eq. (\ref{Adpp}), where $P_{k,0}$ represents the oscillation period of the reference model without any perturbations.

\begin{figure*}
    \centering
	\includegraphics[width=0.32\columnwidth]{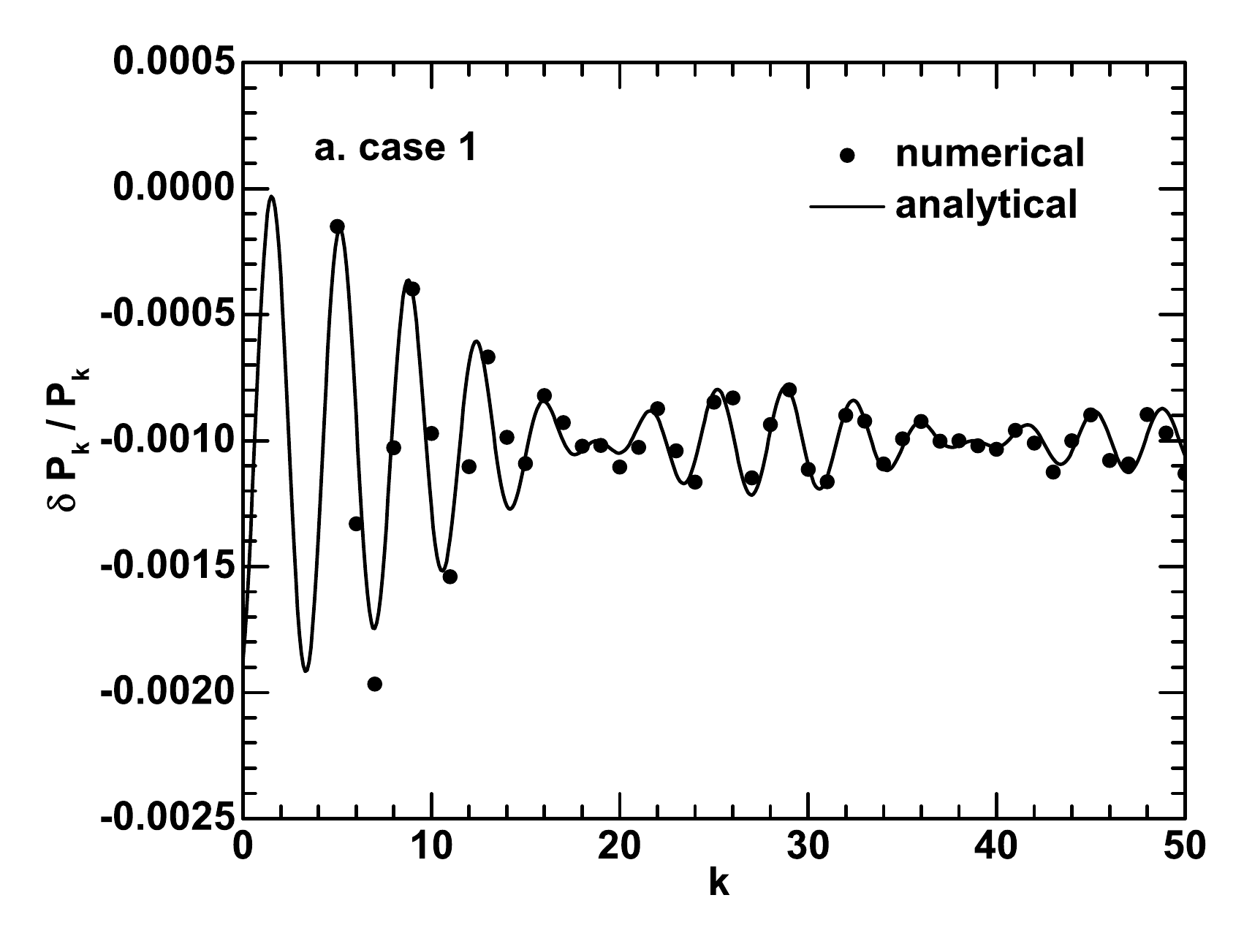}
	\includegraphics[width=0.32\columnwidth]{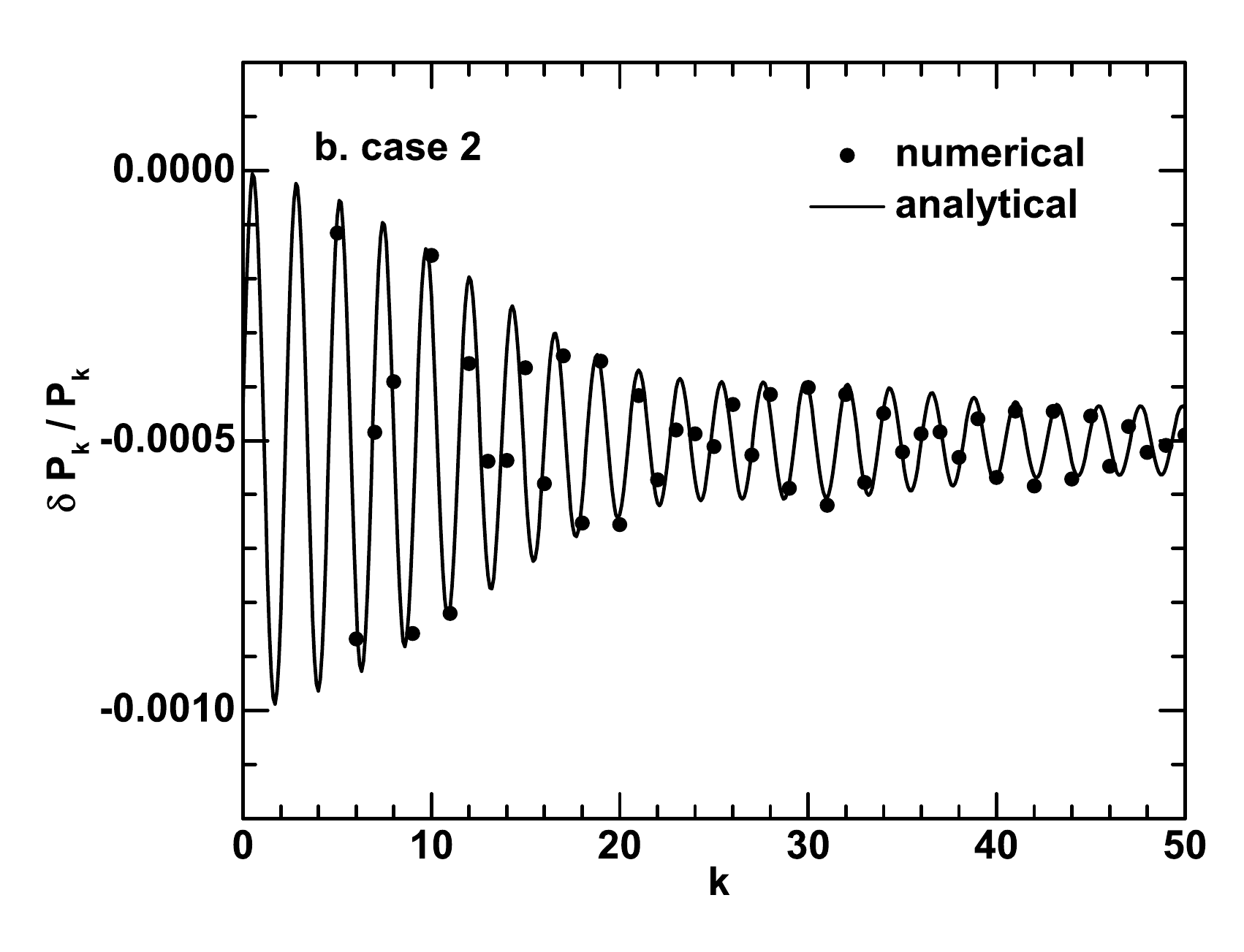}
	\includegraphics[width=0.32\columnwidth]{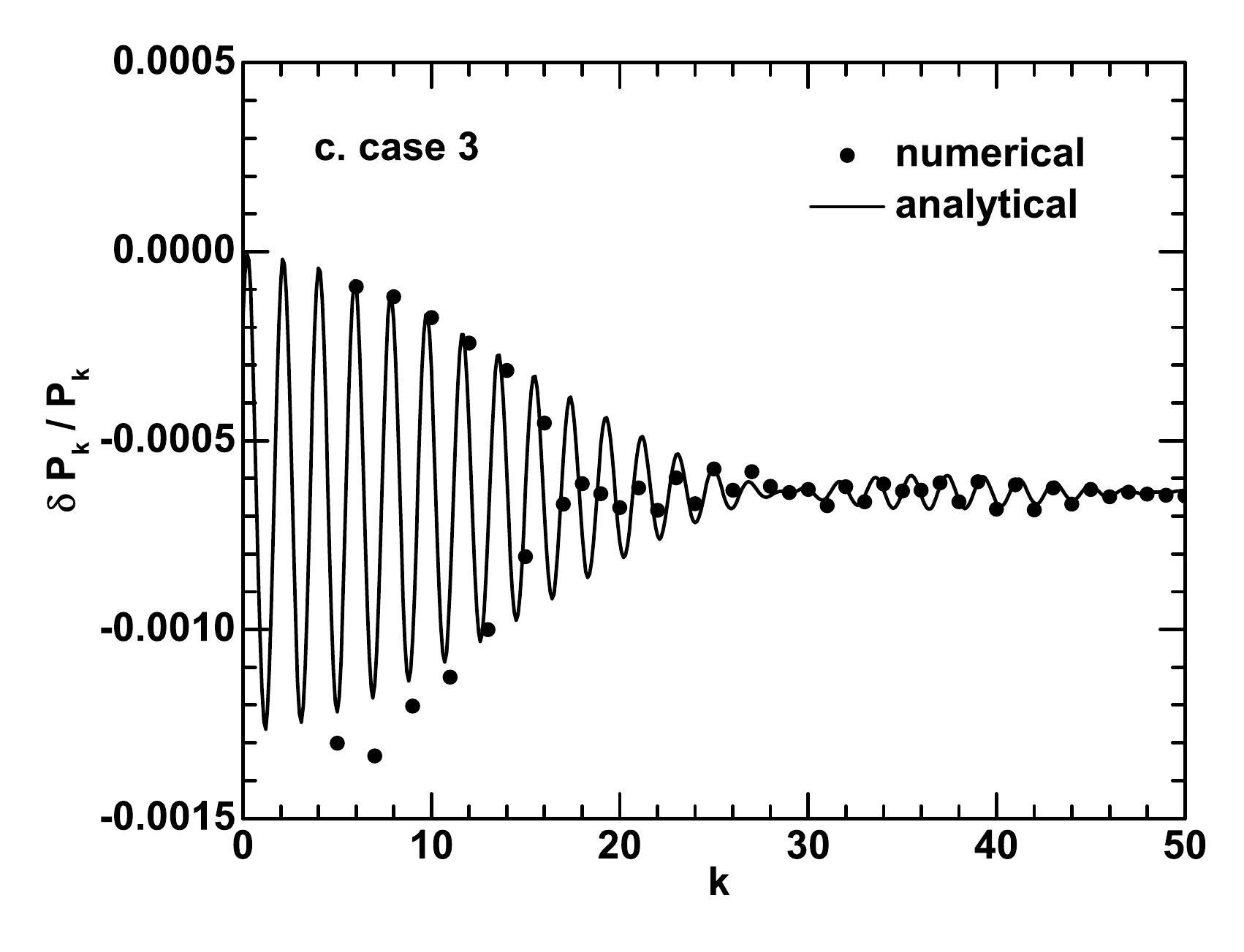}
    \caption{Variations of the oscillation periods with different perturbations of $N$. $k$ is the node of modes, the dots are $P_k/P_{k,0}-1$ calculated by using stellar oscillation code, the lines are calculated by using Eq.(\ref{zqsdPP}). In panel a, $z = \delta N/N = 0.02 $ in $0.25 \leq u \leq 0.3 $. In panel a, $z = 0.02 [(u-0.4)/0.05] $ in $0.4 \leq u \leq 0.45 $. In panel c, $z = 0.02 \sin [ \pi (u-0.5)/0.05] $ in $0.5 \leq u \leq 0.55 $.}
    \label{AfigdPP}
\end{figure*}

In case 1, the perturbation of $N$ is
\begin{eqnarray} \label{Acase1z}
z(u) = \left\{ {\begin{array}{*{20}{c}}
   {0, u < {u_1} }  \\
   {A, {u_1} \le u \le {u_2}}  \\
   {0, u > {u_2} }  \\
\end{array}} \right..
\end{eqnarray}%
Based on Eq.(\ref{Adpp}), the resulting variation of oscillation period is
\begin{eqnarray} \label{Acase1zqsdPP}
\frac{{\delta {P_k}}}{{{P_k}}} \approx \frac{A}{{2 \pi (k + {k_0})}}\{ \cos [2\pi (k + {k_0}){u_2}] - \cos [2\pi (k + {k_0}){u_1}]\} - A(u_2-u_1).
\end{eqnarray}%
We have set $u_1=0.25$, $u_2=0.3$ and $A=0.02$. The results are shown in Fig.\ref{AfigdPP}a.

In case 2, the perturbation of $N$ is
\begin{eqnarray} \label{Acase2z}
z(u) = \left\{ {\begin{array}{*{20}{c}}
   {0, u < {u_1} }  \\
   {A\frac{{u - {u_1}}}{{{u_2} - {u_1}}}, {u_1} \le u \le {u_2} }  \\
   {0, u > {u_2} }  \\
\end{array}} \right.
\end{eqnarray}%
Based on Eq.(\ref{Adpp}), the resulting variation of oscillation period is
\begin{eqnarray} \label{Acase2zqsdPP}
&&\frac{{\delta {P_k}}}{{{P_k}}} \approx \frac{A}{{2 \pi (k + {k_0})}}\{ \cos [2\pi (k + {k_0}){u_2}] \\ \nonumber
&& - \frac{{\sin [2\pi (k + {k_0}){u_2}] - \sin [2\pi (k + {k_0}){u_1}]}}{{2\pi (k + {k_0})({u_2} - {u_1})}}\} - \frac{1}{2}A({u_2} - {u_1})
\end{eqnarray}%
We have set $u_1=0.4$, $u_2=0.45$ and $A=0.02$. The results are shown in Fig.\ref{AfigdPP}b.

In case 3, the perturbation of $N$ is
\begin{eqnarray} \label{Acase3z}
z(u) = \left\{ {\begin{array}{*{20}{c}}
   {0,u < {u_1}}  \\
   {A\sin (\frac{{u - {u_1}}}{{{u_2} - {u_1}}}\pi ),{u_1} \le u \le {u_2}}  \\
   {0,u > {u_2}}  \\
\end{array}} \right.
\end{eqnarray}%
Based on Eq.(\ref{Adpp}), the resulting variation of oscillation period is
\begin{eqnarray} \label{Acase3zqsdPP}
&&\frac{{\delta {P_k}}}{{{P_k}}} \approx  - \frac{{{A_n}({u_2} - {u_1})}}{\pi }\{ 2 + \frac{{\sin [2\pi (k + {k_0}){u_1}] + \sin [2\pi (k + {k_0}){u_2}]}}{{1 - {{[2(k + {k_0})({u_2} - {u_1})]}^2}}}\}
\end{eqnarray}%
We have set $u_1=0.5$, $u_2=0.55$ and $A=0.02$. The results are shown in Fig.\ref{AfigdPP}c.

It can be found that Eq. (\ref{Adpp}) well reproduces the variations of the oscillation periods in all cases of perturbations of $N$. The differences between the numerical results and the analytical results are relatively significant in the low wave number region, while they are small in the high wave number region. This discrepancy is likely due to the limitations of the JWKB approximations and the asymptotic expression of $g$-mode periods, which may not hold well in the low wave number region.

\end{document}